\documentclass[a4paper,11pt]{article}
\pdfoutput=1 

\usepackage{jcappub} 

\usepackage[T1]{fontenc} 

\usepackage{xcolor}
\usepackage{wrapfig}
\usepackage{multirow}
\usepackage{caption}
\usepackage{subcaption}
\usepackage{sidecap}
\usepackage{float}
\usepackage{bm}
\usepackage[utf8]{inputenc}
\graphicspath{ {images/} }

\usepackage{appendix}

\def\code#1{\texttt{#1}} 

\title{Cosmological measurements from the CMB and BAO are insensitive to the tail probability in the assumed likelihood}

\author[a,b,c,1]{J. Krywonos,\note{Corresponding author.}}
\author[b,d]{S. Paradiso,}
\author[b,d,a,*]{A. Krolewski,}
\author[b,d,e]{S.~Joudaki,}
\author[b,d,a]{W.J. Percival}

\affiliation[a]{Perimeter Institute for Theoretical Physics, 31 Caroline St North, Waterloo, ON N2L 2Y5, Canada}
\affiliation[b]{Waterloo Centre for Astrophysics, University of Waterloo, Waterloo, ON N2L 3G1, Canada}
\affiliation[c]{Department of Physics and Astronomy, York University, Toronto, ON M3J 1P3, Canada}
\affiliation[d]{Department of Physics and Astronomy, University of Waterloo, Waterloo, ON N2L 3G1, Canada}
\affiliation[e]{Institute of Cosmology \& Gravitation, Dennis Sciama Building, University of Portsmouth,
Portsmouth, PO1 3FX, United Kingdom}
\affiliation[*]{CITA National Fellow}

\emailAdd{jkrywonos@perimeterinstitute.ca}

\abstract{When fitting cosmological models to data, a Bayesian framework is commonly used, requiring assumptions on the form of the likelihood and model prior.
In light of current tensions between different data, it is interesting to investigate the robustness of cosmological measurements to statistical assumptions about the likelihood distribution from which the data was drawn. We consider the impact of changes to the likelihood caused by uncertainties due to the finite number of mock catalogs used to estimate the covariance matrix, leading to the replacement of the standard Gaussian likelihood with a multivariate $t$-distribution. These changes to the likelihood have a negligible impact on recent cosmic microwave background (CMB) lensing and baryon acoustic oscillation (BAO) measurements, for which covariance matrices were measured from mock catalogs.
We then extend our analysis to perform a sensitivity test on the Gaussian likelihoods typically adopted, considering how increasing the size of the tails of the likelihood (again using a $t$-distribution) affects cosmological inferences. For an open $\Lambda$CDM model constrained by BAO alone, we find that increasing the weight in the tails shifts and broadens the resulting posterior on the parameters, with a $\sim$0.2--$0.4\sigma$ effect on $\Omega_\Lambda$ and $\Omega_\mathrm{k}$. In contrast, the CMB temperature and polarization constraints in $\Lambda$CDM showed less than $0.03\sigma$ changes in the parameters, except for $\{\tau$, ln(\(10^{10}A_\mathrm{s})\), $\sigma_8$, $S_8$, $\sigma_8\Omega_\mathrm{m}^{0.25}$, $z_\mathrm{re}$, $10^9A_\mathrm{s}e^{-2\tau}\}$ which shifted by around $0.1$--$0.2\sigma$. 
If we use solely $\ell < 30$ data, the amplitude $A_\mathrm{s} e^{-2\tau}$ varies in the posterior mean by $0.7\sigma$ and the error bars increase by 6\%. We conclude, at least for current-generation CMB and BAO measurements, that uncertainties in the shape and tails of the likelihood do not contribute to current tensions.}

\begin{document}
\maketitle
\flushbottom



\section{Introduction}\label{intro}

The standard cosmological model with a cosmological constant ($\Lambda$) and cold dark matter (CDM) model agrees well with an extensive range of observational results. Experiments that support this $\Lambda$CDM model include spectroscopic galaxy surveys such as the extended Baryon Oscillation Spectroscopic Survey (BOSS and eBOSS; \cite{BOSS, eBOSS}), photometric weak lensing surveys such as the Dark Energy Survey (DES; \cite{DES2021CosmoConstraints}), cosmic microwave background (CMB) measurements like Planck \cite{Planck2018cosmoparams}, and the Pantheon and Union supernova compilations \cite{Pantheon2022, Union2023}. However, there are tensions in the derived parameters of the $\Lambda$CDM model from different datasets. In particular, the Hubble constant ($H_0$) measurements from local distance ladder measurements \cite{Hubbletension} disagree with those from the CMB \cite{Planck2018cosmoparams} or from Big Bang nucleosynthesis (BBN) analysed together with Baryon Acoustic Oscillation (BAO) data \cite{SDSSeBOSS}. In addition, different measurements of the clustering amplitude of matter ($S_8$) disagree, particularly those from weak lensing \cite{KiDS2021, HSC:Li2023} when compared to the predictions from \textit{Planck}.\footnote{This discrepancy between weak lensing and CMB measurements is not straightforward. For example, there exists a recent combined analysis of the lensing datasets of the Kilo-Degree Survey (KiDS)+DES  that shifts the $S_8$ posterior towards values favored by \textit{Planck} \cite{KiDSandDES:2023}.} 

Given these tensions and future improvements in data precision \cite{desi2019,euclid}, there has been a renewed focus on parameter uncertainties stemming from systematic errors. For example, the reionization optical depth, $\tau$, as measured using the CMB is very sensitive to changes in the analysis pipeline. The value has changed significantly between analyses: the WMAP experiment measured $\tau = 0.17\pm0.04$ \cite{WMAP:2003} from the year one data, and $\tau = 0.089\pm0.014$ after nine years of data \cite{WMAP:2013}. In contrast, \textit{Planck} published $\tau = 0.097\pm0.038$ in its first set of results \cite{Planck:2013pxb}, and $\tau = 0.058\pm0.006$ using the final data and an updated pipeline \cite{Tristram:2020wbi}. Research collaborations independent from \textit{Planck}, such as \code{BeyondPlanck} \cite{Paradiso2022}, continue to improve the analysis technique to achieve a higher control over the uncertainty propagation and provide a more robust estimate.

In the standard Bayesian analysis pipelines used to make inferences from cosmological data (e.g.\ \cite{Cobaya}), credible intervals for cosmological model parameters are determined by exploring the posterior surface.\footnote{For a comparison of Bayesian credible intervals to frequentist confidence regions, see \cite{percivalmatching}.} The results depend on the prior assumed for each parameter and any assumptions made when determining the likelihood. The likelihood is typically approximated to be Gaussian, but this is generally not exact. For example, the likelihood of CMB multipoles $C_\ell$ is a Wishart distribution, which deviates from Gaussianity particularly at low $\ell$ where the central limit theorem does not hold \cite{Percival2006,HamimecheLewis2008}. A further complication results from errors in the covariance matrix \cite{DodelsonSchneider2013,Taylor2013}, which also modify the simple Gaussian likelihood.

In this paper, we consider how assumptions about the form of the likelihood affect recovered credible intervals for fits to the Sloan Digital Sky Survey (SDSS) \cite{SDSSeBOSS} and \textit{Planck} \cite{Planck2018cosmoparams} data. We first consider the case where the covariance on an intermediate statistic itself has an error. This is the situation when covariance matrices for correlation functions or power spectra are determined from sets of mocks, as for recent analyses of eBOSS data \cite{SDSSeBOSS} and CMB lensing \cite{Plancklensing} data from \textit{Planck}, for example. In this case, to complete a Bayesian analysis we need a prior on the covariance matrix, and a number of suggestions have previously been made for this, including an Independence-Jeffreys prior \cite{SellentinHeavens}, or a prior matching frequentist confidence intervals with Bayesian credible intervals \cite{percivalmatching}. We consider the effect of this choice of prior, comparing against the situation where we ignore the errors in the covariance matrix.

We then extend this work to test the sensitivity to the Gaussian assumptions for the likelihood when fitting BAO and CMB data.
This can be considered a sensitivity analysis for the cosmological inferences made. Within the fields of biostatistics and epidemiology, sensitivity analyses play a crucial role in assessing the robustness of conclusions drawn from observational data (eg. \cite{sensitivityanalysis:epidemoilogical, sensitivityanalysis:review, sensitivityanalysis:principle}). They are a critical way of assessing the impact, effect, or influence of key modelling choices like definitions of outcomes, protocol deviations, missing data assumptions, outliers, or prior specifications. Many experiments make a Gaussian approximation for the likelihood even if the model does not intrinsically support this, or the central limit theorem does not hold. Additionally, noise is not necessarily Gaussian distributed, and large-scale structure analyses often use quasi-linear modes where gravitational non-Gaussianity is non-negligible.
Hence, considering heavy-tailed likelihoods in cosmology is well-motivated and this has been done in past work
\cite{Chen2003, Bailey2017, Feeney2018}. We focus on the tails of the distribution assumed for the likelihood, using the multivariate \textit{t}-distribution as a replacement for the Gaussian distribution to vary the fraction of probability within the tails and determine how sensitive the final parameter constraints are to this choice. 

Section~(\ref{methods}) provides background information on statistical techniques and introduces the procedure we implemented to stress test components of the data analysis. Then, Section~(\ref{sec:data}) outlines the datasets that we studied. Following this, Section~(\ref{results1}) considers analyses where the covariance matrix itself has errors, while Section~(\ref{results2}) details our sensitivity analysis of more general BAO and CMB measurements. Finally, we conclude in Section~(\ref{conclusion}). 

\section{Statistical Techniques}\label{methods}

We will focus on Bayesian parameter inference (e.g. \cite{Bayesiantextbook}), where one derives credible regions for parameters based on the posterior, inferred using Bayes' theorem:
\begin{equation}\label{proptoBayes}
    \mathcal{P}(\mathrm{H} | \mathrm{D}) \propto \mathcal{P}(\mathrm{H}) \mathcal{P}(\mathrm{D} | \mathrm{H})\,,
\end{equation}
where \(H\) is the hypothesis to be tested and \(D\) is the data. The posterior \(\mathcal{P}(\mathrm{H} | \mathrm{D})\) is determined from the likelihood \(\mathcal{P}(\mathrm{D} | \mathrm{H})\) and prior \(\mathcal{P}(\mathrm{H})\). 
So, to determine the posterior distribution of the model parameters, the prior must be chosen to reflect the knowledge of the parameters before the new data is considered, and a likelihood function specified. The multivariate Gaussian (Normal) distribution is often adopted for the likelihood, if the distribution from which the data are drawn is not known: \cite{Trotta:2017}
\begin{equation}\label{likelihood}
    \mathcal{P}(\mathbf{x}|\mathbf{p}, \mathbf{\Sigma}) = \frac{1}{\sqrt{|2\pi\mathbf{\Sigma}|}}\mathrm{exp}\Big[-\frac{1}{2}\chi^2(\mathbf{x},\mathbf{p},\mathbf{\Sigma}^{-1})\Big].
\end{equation}
Here, \(\mathbf{\Sigma}\) is the covariance matrix, the data is \(\mathbf{x}^d\), the data model is \(\mathbf{x}(\mathbf{p})\), the parameters are \(\mathbf{p}\), and 
\begin{equation}
    \chi^2(\mathbf{x}, \mathbf{p}, \mathbf{\Sigma}^{-1}) \equiv \sum_{ij}[x_i^d - x_i(\mathbf{p})]\Sigma_{ij}^{-1}[x_j^d - x_j(\mathbf{p})].
\end{equation}

In order to test how robust modern cosmology is to the choices made for the likelihood, we consider two different cases as detailed in the next section.

\subsection{Covariance Matrix Estimated From Simulations}  \label{sec:intro_sim_cov}

An unbiased estimate of the true covariance matrix \(\Sigma\) from a set of simulated data is given by
\begin{equation}
    S = \frac{1}{n_s-1}\sum_{i=1}^{n_s}(x^m_i-\Bar{x}^m)(x^m_i-\Bar{x}^m)^T,
\end{equation}
where $n_s$ is the number of mock simulations, \(x^m_i\) are the mock data, and \(\Bar{x}^m\) is the mean of \(x^m_i\). Inverting this estimate ($S^{-1}$), however, gives a biased estimate of the inverse covariance $\Sigma^{-1}$ \cite{hartlap2007}. This systematically biases the credible regions placed on cosmological parameters. Simply correcting the skewness in the inverse covariance matrix as advocated by \cite{hartlap2007} does not correct the credible regions because it does not take into account the way in which the posterior surface is explored, which can be considered a constrained inversion of the inverse covariance back to a covariance of new parameters \cite{percivalmatching}. An approximate correction for this effect is to preserve the Gaussian likelihood function and simply scale the covariance \cite{Percival2013,percivalmatching}, such that the matrix to invert is $S'$ where
\begin{eqnarray}  \label{eq:Sprime}
    S'&=&\frac{(n_s-1)[1+B(n_d-n_p)]}{n_s-n_d+n_p-1}S\,,\\
    B&=&\frac{(n_s-n_d-2)}{(n_s-n_d-1)(n_s-n_d-4)}\,.
\end{eqnarray}
Here $n_d$ is the number of data points and $n_p$ is the number of parameters. 

Ref.~\cite{SellentinHeavens} instead considered a fully Bayesian approach where an Independence-Jeffreys prior is placed on the covariance matrix, and it is considered a random variable in the analysis. This results in a multivariate \textit{t}-distribution posterior \cite{SellentinHeavens},
\begin{equation}\label{eq: SellHeavtdist}
    \mathcal{P}(\bm{\mu}|\mathbf{x}_0, S) \propto \left[1+\frac{(\mathbf{x}_0-\bm{\mu})S^{-1}(\mathbf{x}_0-\bm{\mu})^T}{n_s -1}\right]^{-\frac{m}{2}},
\end{equation}
where \(\mathbf{x}_0\) are the data with dimension \(n_d\). For the Independence-Jeffreys prior we have $m=n_s$. Ref.~\cite{percivalmatching} considered instead a prior on the covariance matrix designed to match Bayesian credible intervals with frequentist confidence intervals. Again, this results in a multivariate \textit{t}-distribution posterior, but this time with
\begin{equation}
    m = n_p+2+\frac{n_s-1+B(n_d-n_p)}{1+B(n_d-n_p)}\,.
\end{equation}
We note that for the multivariate \textit{t}-distribution, $S$ is referred to as the scale matrix, as the covariance is $(m-n_d)S/(m-n_d-2)$.

Using either the Independence-Jeffreys or frequentist matching prior for the true covariance matrix results in a posterior with larger tails than a Gaussian, increasing the likelihood of parameter values further away from the central region. This raises the question of how much this changes the cosmological constraints for previously published datasets. We will investigate this for both choices of priors, hereafter calling the posteriors the Independence-Jeffreys \textit{t}-distribution and the matching prior \textit{t}-distribution, comparing against a Gaussian posterior where the error in the covariance matrix is ignored.
 
 \subsection{Sensitivity Analysis}  \label{sec:method-sensitivity}

In the previous section, we considered well motivated changes within a Bayesian context to the likelihood following consideration of how the covariance matrix was estimated. Changes to the likelihood can also form part of a sensitivity analysis, where we consider how robust our inferences are to the analysis method assumed. The approximation of a particular form for the likelihood may not be easy to determine, but we can easily test how important that approximation is for particular parameters. For this, following on from Section~\ref{sec:intro_sim_cov}, we consider the general form of the multivariate \textit{t}-distribution:
\begin{equation}\label{eq:multi-tdist}
    f(\mathbf{x}) = \frac{\Gamma[(\nu + n_d)/2]}{\Gamma(\nu/2) \nu^{n_d/2} \pi^{n_d/2} \mid\mathbf{\Sigma}\mid^{1/2}} \Big[1+\frac{1}{\nu}(\mathbf{x}-\bm{\mu})^T \mathbf{\Sigma}^{-1}(\mathbf{x}-\bm{\mu})\Big]^{-(\nu+n_d)/2}.
\end{equation}
Here, $\mathbf{x}$ is the data and $\bm{\mu}$ is the model, which both have dimension $n_d\times 1$. $\mathbf{\Sigma}$ is the $n_d\times n_d$ scale matrix and $\nu$ is the number of degrees of freedom. In the limit $\nu\to\infty$, the distribution tends towards a Gaussian form. Using the multivariate \textit{t}-distribution we can set the number of degrees of freedom based on how much more probability is in the tails of the chosen \textit{t}-distribution compared to a Gaussian. 

\begin{figure}
    \centering
    \includegraphics[width=0.5\linewidth]{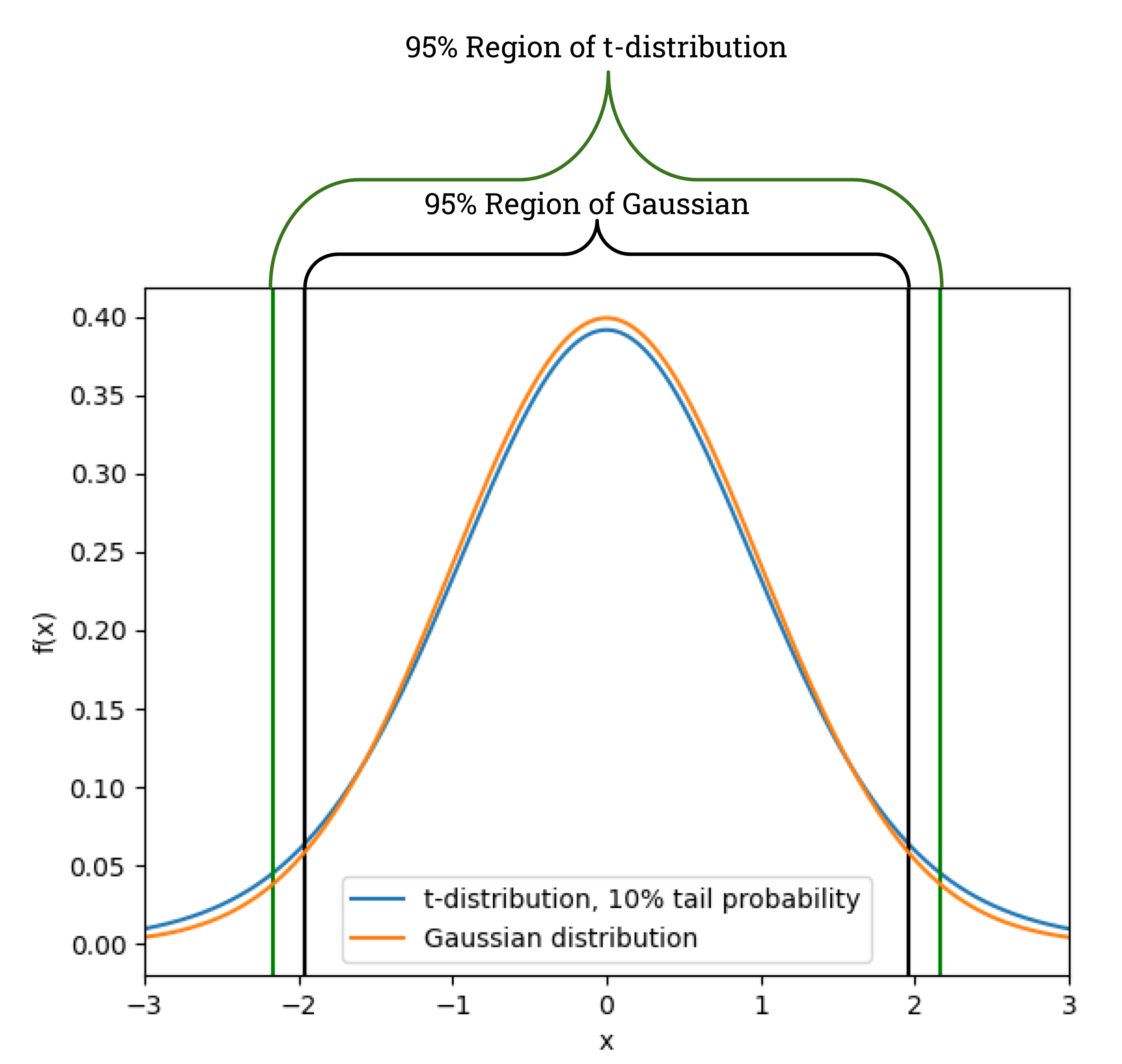}
    \caption{A Gaussian distribution (orange) versus the Student \textit{t}-distribution with its $2\sigma$ interval 10\% larger than the Gaussian's (blue).}
    \label{fig:gaussvstdist}
\end{figure}

In order to quantify the extent to which we are adjusting the likelihoods, we consider the multivariate \textit{t}-distribution for a single data point, marginalized over the other data points, and calculate the $2\sigma$ interval for different values of $\nu$. We choose the two $\nu$ values which correspond to $2\sigma$ intervals that are 1\% or 10\% larger than those of a marginalized multivariate Gaussian with matching variance. To do this, we use the marginalized multivariate \textit{t}-distribution, which is the Student \textit{t}-distribution,
\begin{equation}
    f(\mathbf{x}) = \frac{\Gamma[(\nu+1)/2]}{\Gamma(\nu/2)\sqrt{\nu\pi}} \Bigg[1+\frac{x^2}{\nu}\Bigg]^{-(\nu+1)/2}\,,
\end{equation}
and calculate the cumulative distribution function and evaluate the 2.5\textsuperscript{th} and 97.5\textsuperscript{th} percentiles. These are then compared to the percentiles of the Gaussian distribution so the value of $\nu$ that increased the region by 1\% and 10\% could be determined. For the 1\% scaling, $\nu=104$, and for 10\% scaling, $\nu=13$. An example of how the distribution changes is shown in Fig.~(\ref{fig:gaussvstdist}) for the 10\% case. In the Monte Carlo code, the multivariate Gaussian distribution was replaced with Eq.~(\ref{eq:multi-tdist}) using the derived values of $\nu$, along with the corresponding value of $n_d$ for the dataset considered.

\section{Data} \label{sec:data}
\subsection{SDSS BAO}

Over 20 years, the Sloan Digitial Sky Survey (SDSS; \cite{York:2000}) has undertaken a series of galaxy redshift surveys, from which the BAO scale can be measured. At low redshift, $0.07<z<0.2$, there is the Main Galaxy Sample (MGS;  \cite{Howlett:2015,Ross:2015}) from Data Release 7 (DR7; \cite{Abazajian:2009}); at $0.2 < z < 0.5$, the Baryon Oscillation Spectroscopic Survey (BOSS; \cite{Dawson:2013}) from DR12 (\cite{Alam-DR11&12:2015}); and at $z > 0.6$,
extended BOSS (eBOSS) \cite{Dawson:2016} from DR16 (\cite{Ahumada:2020}). 
In our work, we focus on the analyses where galaxies are used as discrete tracers of the density field, and the BAO were measured and fitted from these samples using both the correlation function and power spectrum. For the 2-point functions, covariance matrices are calculated using mock catalogues. 

We primarily focus on fitting the DR12 BOSS LRG correlation function \cite{RossBOSS} for redshift bins $z=0.38$ and $z=0.61$ (since incorporating bin $z=0.51$ adds negligible extra information), as these data represent the best large-scale structure data currently available. 
We directly fit the cosmological parameters to the two-point correlation function using the post-reconstruction damped BAO model of \cite{Seo16}.
In Appendix~(\ref{app:BAOcompressed}), we instead work with all BAO samples described above (which are summarized in Table~(\ref{table:covariances}), including the number of mocks, size of the data vector, and number of parameters in the model), but use their publicly released compressed parameters ($\alpha_{\parallel}$, $\alpha_{\perp}$) rather than re-fitting the correlation function measurements. 
The SDSS team conducted these fits in two phases, first compressing the correlation function into (model-independent) Alcock-Paczysnki (AP) parameters, $\alpha_{\parallel}$ and $\alpha_{\perp}$, which parameterise the measured dilation of the BAO peak along and across the line of sight with respect to a fiducial cosmology. They then fit cosmological parameters to the compressed dataset. Our likelihood directly fits the cosmological parameters to the correlation functions by converting them into $\alpha_{\parallel}$ and $\alpha_{\perp}$ and then shifting the post-reconstruction BAO model of \cite{RossBOSS} by these AP parameters.
The likelihood of \cite{RossBOSS} includes a prior on the bias $B_0$; we first split this prior from the likelihood, and then add it back after scaling $\chi^2$ or changing from a Gaussian to a $t$-distribution.

We use the publicly available post-reconstruction correlation functions and covariance matrices of \cite{RossBOSS},\footnote{Data available at \url{https://github.com/ashleyjross/BAOfit/tree/master/exampledata/Ross_2016_COMBINEDDR12} and our code was adapted from their likelihood in \url{https://github.com/ashleyjross/LSSanalysis/tree/main}.}
and removed the Hartlap factor before switching to either a frequentist-matching prior or Independence-Jeffreys \textit{t}-distribution. We considered the open \(\Lambda\)CDM (o\(\Lambda\)CDM) model that allows for non-zero curvature, sampling over cosmological parameters $\Omega_\mathrm{m}$, and $\Omega_\mathrm{k}$ and fixing $\Omega_\mathrm{b} = 0.0468$ and  $H_0 = 70$. We use \code{Cobaya} \cite{Cobaya} to obtain the cosmological parameter constraints and considered the chains converged for Gelman-Rubin $R-1 \leq 0.02$ \cite{Gelman92} along with a further exploration of the tails beyond the 95\% confidence interval, with a permitted quantile chain variance of 0.02 standard deviations in the 95\% confidence interval. 

\subsection{\textit{Planck} CMB Lensing}

We also consider the \textit{Planck} 2018 CMB lensing measurements  \cite{Plancklensing}. Gravitational lensing creates distinctive, non-Gaussian structure in CMB temperature and polarization maps, which can be extracted using  quadratic estimators \citep{HuOkamoto02}. The \textit{Planck} team applied these estimators to the \textit{Planck} temperature ($T$) and $E$-mode polarization ($E$) maps to produce a map of the CMB lensing potential $\hat{\phi}$ across 60\% of the sky. Estimating the lensing potential power spectrum requires (i) a mean field normalization correction, (ii) a subtraction of noise biases (arising from the disconnected four-point function of the Gaussian CMB, non-primary couplings of the connected four-point function, and point source biases), and (iii) the application of a simulation-determined Monte Carlo calculation. Complications such as the Galactic mask and sky varying noise couple previously independent Fourier modes of $\phi$. The covariance matrix used to include such effects in \cite{Plancklensing} was estimated by applying lensing reconstruction to 240 realistic FFP10 CMB simulations.\footnote{It was confirmed in private communication with Julien Carron that the number of simulations should be 240, not the 300 mentioned in \cite{Plancklensing}.} Any dependence on the theoretical model of the CMB power spectra is removed by marginalization over the primary CMB power spectrum, which adds a term to the covariance matrix (Eq.~(34) in \cite{Plancklensing}). It is this covariance matrix containing the additive correction that we include in our analysis.

We test the sensitivity of the measurements from CMB lensing for the \(\Lambda\)CDM model, sampling over the cosmological parameters $\Omega_\mathrm{b}h^2$, $\Omega_\mathrm{c}h^2$, $100\theta_\mathrm{MC}$, ln$(10^{10}A_\mathrm{s})$ and $n_s$. We used a modified version of the code \code{CosmoMC} \cite{CosmoMC} to obtain the parameter constraints with a convergence criterion of $R-1 \leq 0.01$ and further sample the tails beyond the 99\% confidence level, with a limit on the quantile chain variance of 0.2 standard deviations. To calculate the covariance matrix, the number of simulations used was \(n_s = 240\), the data vector had \(n_d = 9\) (corresponding to the conservative multipole range $8 \leq \ell \leq 400$), and the number of parameters was \(n_p = 5\) \cite{Plancklensing}. We also checked that using the aggressive multipole range $8 \leq \ell \leq 2048$ gave results similar to the conservative range, so for this test the data vector was \(n_d = 14\).\footnote{In private communication with Julien Carron we learned that \(n_d = 14\) not 16 as indicated in \cite{Plancklensing}.} Unlike SDSS, only the Hartlap correction factor \cite{hartlap2007} was considered, so we removed this and then changed from the Gaussian to the Independence-Jeffreys and matching prior \textit{t}-distribution posteriors. We used the lensing convergence power spectrum from the minimum variance (MV) combination of $T$ and $E$ maps.\footnote{The .inputparams files used in 
\code{CosmoMC}, obtained from \url{https://pla.esac.esa.int/##cosmology}, are called `base\_lensing\_lenspriors.inputparams' for the conservative range and `base\_lensing\_lenspriors\_pttagr2.inputparams' for the aggressive range.}.

\subsection{\textit{Planck} CMB Temperature and Polarization}

Next, we consider the \textit{Planck} CMB measurements of temperature and polarization ($T$\&$P$) \cite{Planckresults}. \textit{Planck} considers different combinations of auto- and cross-correlations of the $T$ and $E$ spectra, as well as different $\ell$ ranges. These are contained in separate codes exploring high-$\ell$ and low-$\ell$ multipoles separately \cite{Planck2018-05}. In detail, \code{Plik} is the joint $TT$, $EE$ and $TE$ likelihood in the multipole range $30$--$2508$ for $TT$, and $30-1996$ for $TE$ and $EE$. It is based on the binned cross-spectra from 100, 145 and 217 GHz channels, and represents the likelihood as a correlated Gaussian:
\begin{equation}
-\ln{\mathcal{L}(\hat{\bf{C}}|\bf{C(\theta)})} = \frac{1}{2}
\left[\left(\hat{\bf{C}}-\bf{C(\theta)}\right)^T\Sigma^{-1}
\left(\hat{\bf{C}}-\bf{C(\theta)}\right)\right] + \text{constant}\,,
\end{equation}
where $\hat{\bf{C}}$ is the vector with observed spectra, $\bf{C}(\theta)$ are the predicted spectra for the cosmological parameter set $\theta$, and $\Sigma$ is the covariance matrix as computed for a fiducial realisation. Even though the Gaussian shape is an approximation to the true Wishart distribution, it has been demonstrated to perform reasonably well even for $\ell\sim 30$, as discussed in \cite{HamimecheLewis2008}.
In the \cite{Planck2018} analysis,
large angular scales ($\ell < 30$) use the  \code{SimAll} ($EE$) and \code{commander} ($TT$) likelihoods. The latter is based on a Gaussianised Blackwell-Rao estimator of the $TT$ power spectrum from foreground cleaned CMB samples \cite{GBR}, whereas the former consists of a brute-force inversion of the likelihood for polarisation power spectra estimates from foreground-cleaned maps \cite{Planck2018-05}. Hence \code{commander} is a Gaussian likelihood, and \code{SimAll} is not.
Our procedure consists of changing Gaussian likelihoods to multivariate $t$-distributions,
and should not be applied to likelihood
forms that are already non-Gaussian.

We therefore use the LFI-based likelihood \code{bflike} for large angular scales \cite{Planck2018-05}, replacing \code{commander} and \code{SimAll}.
\code{bflike} is a map-based Gaussian likelihood, and unlike \code{commander},
it includes polarization as well as temperature. The likelihood for \code{bflike} is
\begin{equation}
    \mathcal{L}(C_\ell)=\mathcal{P}(\bf{m}|C_\ell) =
    \frac{1}{\sqrt{2\pi|S(\theta)+N|}}\exp\lbrace-\frac{1}{2}\bf{m}^T(S(\theta)+N)^{-1}\bf{m}\rbrace,
\end{equation}
where $\bf{m}$ is the CMB-plus-Noise map, and $S(\theta)+N$ is the Signal-plus-Noise covariance. The signal covariance here is computed for every cosmological parameter set $\theta$ in order to explore the full joint posterior distribution for $TT$, $TE$, $EE$ and $BB$ power spectra in the multipole range $2-30$. In order for the covariance to be non-singular, it is required that the number of pixels in the Stokes $I$, $Q$, and $U$ maps is larger than $\sim (2\ell_{\rm{max}}+1)^2$. We considered the value of our data vector in the \textit{t}-distribution to be $n_d=2289$ for \code{Plik} and $n_d=6467$ for \code{bflike}, where the latter value comes from the number of unmasked pixels in $I$, $Q$, and $U$ CMB maps at \code{HEALPix} $N_{\mathrm{side}}=16$, and the former from the number of bins used in \textit{TT}, \textit{TE}, and \textit{EE} in different frequency channels (given in Table~(20) of \cite{Planck2018-05}). For the MCMC runs, the base cosmological parameters were {$\Omega_\mathrm{b}h^2$, $\Omega_\mathrm{c}h^2$, $100\theta_\mathrm{MC}$, $\tau$, ln$(10^{10}A_\mathrm{s})$ and $n_s$} when we considered both \code{Plik} ($TTTEEE$) and \code{bflike} (low$TEB$)\footnote{The .ini files we used in \code{CosmoMC} were `batch3/plik\_rd12\_HM\_v22\_TTTEEE.ini' for $TTTEEE$ and `batch2/lowTEB.ini' for low$TEB$.}. For runs with only \code{bflike}, we sampled over cosmological parameters {$\tau$ and ln$(10^{10}A_\mathrm{s})$}. We again used \code{CosmoMC} with the same convergence settings as given above for CMB lensing.

\section{Results for the Likelihood Choice Using Mock-based Covariances}\label{results1}

\begin{figure}[h!]
     \centering
     \begin{subfigure}[b]{0.49\textwidth}
         \centering
         \includegraphics[width=\linewidth]{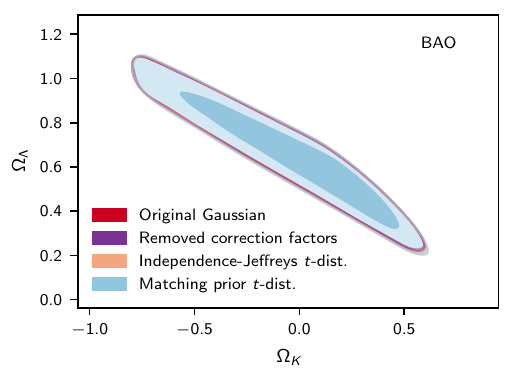}
     \end{subfigure}
     \hfill
     \begin{subfigure}[b]{0.49\textwidth}
         \centering
         \includegraphics[width=\linewidth]{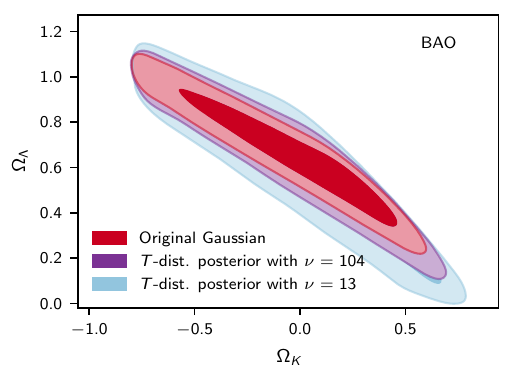}
     \end{subfigure}
     \caption{Comparison of the  \(\Omega_\mathrm{k}\)--\(\Omega_\Lambda\) constraints at 68\% and 95\% credible intervals for the o\(\Lambda\)CDM model using BOSS DR12 LRG correlation function data for redshift bins centered at $z_{\rm eff} = 0.38$ and $0.61$. The left panel shows constraints for the original BAO results from using SDSS's method (red), with the correction factors removed (purple), the matching prior \textit{t}-distribution (orange), and the Independence-Jeffreys \textit{t}-distribution (blue). On the right is again the original Gaussian set-up (red) along with the sensitivity test with 1\% larger tails (purple) and 10\% larger tails (blue).}
    \label{fig:BAOcompare4}
\end{figure}

\begin{table}[h!]
\centering
\begin{tabular}{|c|c|c|}
    \hline
    \multicolumn{3}{|c|}{BOSS DR12 LRG BAO, $z=0.38,0.61$ } \\
    \hline
     & \(\Omega_\Lambda\) & \(\Omega_\mathrm{k}\) \\
   \hline 
    \multirow{2}{*}{Original SDSS method (Gaussian)} & \multirow{2}{*}{\(0.650^{+0.172}_{-0.215}\)} & \multirow{2}{*}{\(-0.035^{+0.424}_{-0.242}\)}\\
     & & \\ 
    \hline
    \multirow{2}{*}{Removed correction factors } & \multirow{2}{*}{\(0.648^{+0.171}_{-0.212}\)} & \multirow{2}{*}{\(-0.036^{+0.417}_{-0.240}\)}\\
     & & \\ 
    \hline
    \multirow{2}{*}{Independence-Jeffreys \textit{t}-dist.} & \multirow{2}{*}{$0.643^{+0.172}_{-0.213}$} & \multirow{2}{*}{$-0.027^{+0.415}_{-0.240}$}\\
     & & \\ 
   \hline
    \multirow{2}{*}{Matching prior \textit{t}-dist.} & \multirow{2}{*}{\(0.642^{+0.172}_{-0.213}\)} & \multirow{2}{*}{\(-0.026^{+0.417}_{-0.240}\)}\\
     & & \\ 
    \hline
\end{tabular}
\caption{Parameter constraints for the o\(\Lambda\)CDM model, using the BOSS DR12 LRG correlation function data for redshift bins centered at $z_{\rm eff} = 0.38$ and $0.61$. The uncertainties are the 68\%  credible intervals. Results were found using the original SDSS method, no correction factors, then the Independence-Jeffreys prior \textit{t}-distribution and the matching prior\textit{ t}-distribution.}
\label{table:results}
\end{table}

We begin by producing cosmological constraints for the o\(\Lambda\)CDM model using SDSS BOSS's original likelihood setup with the Gaussian distribution, before moving on to consider results using instead the \textit{t}-distribution with Independence-Jeffreys or frequentist-matching priors on the covariance matrix. We also compare against the scenario where all of the corrections are removed, equivalent to assuming no error in the covariance matrix. The cosmological constraints are illustrated in Fig.~(\ref{fig:BAOcompare4}), which shows a high level of agreement between the contours. Additionally, the constraints for \(\Omega_\Lambda\) and \(\Omega_\mathrm{k}\) are listed in in Table~(\ref{table:results}). 
The correction factors used by SDSS increase the size of the confidence intervals by 1\%, compared to not using the correction factors and using a Gaussian likelihood.

For the two \textit{t}-distribution likelihoods, the parameters only have up to a $0.04\sigma$ difference in comparison to the original Gaussian results, and the confidence intervals are similar in size to the Gaussian likelihood without the correction factors.  The primary take-home message from Table~(\ref{table:results}) is that none of the choices results in a significant change in constraints. This was also seen when instead of fitting to the correlation function data, we fit to the compressed data (AP parameters) in Appendix~(\ref{app:BAOcompressed}). Therefore, while these methods of marginalizing over the unknown covariance are statistically more rigorous, in practice it does not mean that results from previously published studies need to be reanalyzed. 

\begin{table}[h!]
\centering
\begin{tabular}{|c|c|c|c|c|}
    \hline
    \multicolumn{4}{|c|}{\textit{Planck} CMB Lensing } \\
    \hline
     \multirow{2}{*}{ } & \multirow{2}{*}{\(\sigma_8\Omega_\mathrm{m}^{0.25}\)} & \multirow{2}{*}{\(\sigma_8\)} & \multirow{2}{*}{\(\Omega_\mathrm{m}\)}\\
     & & & \\
   \hline
   \multirow{2}{*}{Original \textit{Planck} 2018 method (Gaussian)} & \multirow{2}{*}{\(0.589 \pm 0.020\)} & \multirow{2}{*}{$0.805^{+0.140}_{-0.074}$} & \multirow{2}{*}{$0.335^{+0.055}_{-0.200}$}\\
     & & & \\
   \hline
   \multirow{2}{*}{Removed correction factors} & \multirow{2}{*}{\(0.590\pm 0.020\)} & \multirow{2}{*}{$0.805^{+0.140}_{-0.076}$} & \multirow{2}{*}{$0.337^{+0.059}_{-0.200}$}\\
     & & & \\
   \hline
   \multirow{2}{*}{Independence-Jeffreys \textit{t}-dist.} & \multirow{2}{*}{\(0.590\pm 0.021\)} & \multirow{2}{*}{$0.803^{+0.140}_{-0.076}$} & \multirow{2}{*}{$0.339^{+0.059}_{-0.200}$}\\
     & & & \\
    \hline
    \multirow{2}{*}{Matching prior \textit{t}-dist.} & \multirow{2}{*}{\(0.590\pm 0.020\)} & \multirow{2}{*}{$0.804^{+0.140}_{-0.075}$} & \multirow{2}{*}{$0.338^{+0.058}_{-0.200}$}\\
     & & & \\
    \hline
\end{tabular}
\caption{Parameter constraints for the \(\Lambda\)CDM model using CMB lensing measurements. The uncertainties are given by the 68\% credible intervals. We considered using either the original Gaussian method, removing the correction factors, or a \textit{t}-distribution with an Independence-Jeffreys or frequentist-matching prior.}
\label{table:LensingResults}
\end{table}

\begin{figure}[h!]
     \centering
      \includegraphics[width=0.6\textwidth]{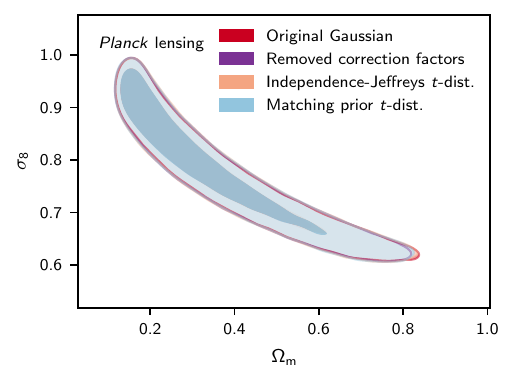}
     \caption{The contour plot for the \(\Lambda\)CDM model, illustrating the \(\Omega_\mathrm{m}\)-\(\sigma_8\) constraints at 68\% and 95\% credible intervals for \textit{Planck} CMB lensing. Shown on the plot is the case of using \textit{Planck}'s original Gaussian method (red), correction factors removed (purple), the Independence-Jeffreys \textit{t}-distribution (orange), and the matching prior \textit{t}-distribution (blue).}
     \label{fig:lensingcompare}
\end{figure}

Next, we consider the \textit{Planck} 2018 CMB lensing measurements within the $\Lambda$CDM model. We first do this for both the original Gaussian setup and then including no correction factors, which are compared in Table~(\ref{table:LensingResults}) and Fig.~(\ref{fig:lensingcompare}) against methods that consider a posterior that allows for the covariance matrix error. The marginalised parameter credible regions are effectively the same with less than a $0.05\sigma$ disagreement. We conclude that the error on the covariance matrix used for the CMB lensing analysis was sufficiently small that the corrections for it are negligible. 

\section{Results from the Sensitivity Analysis}\label{results2}

In the previous section we evaluated the importance of the correction to the likelihood when considering uncertainty in a covariance matrix constructed using mocks. In this section we take this one step further and consider the sensitivity of the analysis to generic changes to the heaviness of the tails, without a specific motivation
from the details of the analysis.
This is a form of sensitivity analysis that is common in other research fields, and is designed to test the robustness of the inferences made to the statistical form assumed for the data (e.g.\ in the presence of ``unknown unknowns'' that may affect the tails of the distribution). For this test, we consider the BAO and CMB lensing data as well as the \textit{Planck} temperature and lensing data. To perform our sensitivity analysis, we have changed the likelihood to a \textit{t}-distribution with degrees of freedom $\nu$ = 104 (1\% extra tail probability) or $\nu$ = 13 (10\% extra tail probability) as described in Section~(\ref{sec:method-sensitivity}).

\subsection{BAO and CMB Lensing}

\begin{table}[h!]
\centering
\begin{tabular}{|c|c|c|}
    \hline
    \multicolumn{3}{|c|}{BOSS DR12 LRG BAO, $z=0.38,0.61$ } \\
    \hline
     & \(\Omega_\Lambda\) & \(\Omega_\mathrm{k}\) \\
   \hline
    \multirow{2}{*}{Original SDSS method (Gaussian)}  & \multirow{2}{*}{\(0.650^{+0.172}_{-0.215}\)} & \multirow{2}{*}{\(-0.035^{+0.424}_{-0.242}\)}\\
     & & \\ 
    \hline 
    \multirow{2}{*}{\textit{T}-dist. likelihood with $\nu$ = 104 (1\% larger tails)} & \multirow{2}{*}{\(0.615^{+0.195}_{-0.221}\)} & \multirow{2}{*}{$0.014^{+0.422}_{-0.241}$}\\ 
     & & \\
    \hline
    \multirow{2}{*}{\textit{T}-dist. likelihood with $\nu$ = 13 (10\% larger tails)} & \multirow{2}{*}{\(0.567\pm 0.241\)} & \multirow{2}{*}{$0.068^{+0.432}_{-0.253}$ }\\ 
     & & \\
    \hline 
\end{tabular}
\caption{The parameter values for the o\(\Lambda\)CDM model using the BOSS DR12 LRG correlation function data for redshift bins centered at $z_{\rm eff} = 0.38$ and $0.61$. We compare the results found using the Gaussian likelihood that SDSS had implemented versus a \textit{t}-distribution likelihood of 1\% or 10\% extra probability in the tails. The uncertainties are given by the 68\% credible intervals.}
\label{table:BAOmargresults}
\end{table}

First, we applied this test to BAO data and we fitted the o\(\Lambda\)CDM model, and focused on \(\Omega_\Lambda\) and \(\Omega_\mathrm{k}\) as the parameters of interest. We have deliberately chosen these data, as we know that the data only weakly constrains these parameters. As shown in Table~(\ref{table:BAOmargresults}) and Fig.~(\ref{fig:BAOcompare4}), the parameter medians increase respectively by (\(\Omega_\Lambda\), \(\Omega_\mathrm{k}\))=($0.2\sigma,0.2\sigma$) and (\(\Omega_\Lambda\), \(\Omega_\mathrm{k}\))=($0.4\sigma,0.3\sigma$) for 1\% and 10\% more power in the tails of the likelihood.
Moreover, the error bars increase by 7\% (25\%) for $\Omega_\Lambda$ and are nearly unchanged for $\Omega_k$.
It is important to note that we find this sensitivity to the heaviness of the tails when constraining an extended model with BAO only. Hence, the increased sensitivity to the likelihood assumptions is likely linked to the data only weakly constraining the model degeneracy we are moving along, which explains why this form of robustness test may be even more relevant to perform on these types of poorly constrained models. 
We find that fitting directly to the correlation function (rather than the intermediate compressed statistics $\alpha_{\parallel}$ and $\alpha_{\perp}$) is critical.
In Appendix~(\ref{app:BAOcompressed}), we fit to $\alpha_{\parallel}$ and  $\alpha_{\perp}$ instead of the correlation function and find that the constraints do not change significantly ($\sim 0.02\sigma$) with the form of the likelihood. 

\begin{table}[h!]
\centering
\begin{tabular}{|c|c|c|c|c|}
    \hline
    \multicolumn{4}{|c|}{\textit{Planck} CMB Lensing } \\
    \hline
     \multirow{2}{*}{ } & \multirow{2}{*}{\(\sigma_8\Omega_\mathrm{m}^{0.25}\)} & \multirow{2}{*}{\(\sigma_8\)} & \multirow{2}{*}{\(\Omega_\mathrm{m}\)}\\
     & & & \\
   \hline
   \multirow{2}{*}{Original \textit{Planck} 2018 method (Gaussian)} & \multirow{2}{*}{\(0.589 \pm 0.020\)} & \multirow{2}{*}{$0.805^{+0.140}_{-0.074}$} & \multirow{2}{*}{$0.335^{+0.055}_{-0.200}$}\\
     & & & \\
   \hline
   \multirow{2}{*}{\textit{T}-dist. likelihood with $\nu$ = 104 (1\% larger tails)} & \multirow{2}{*}{\(0.589\pm 0.020\)} & \multirow{2}{*}{$0.803^{+0.140}_{-0.074}$} & \multirow{2}{*}{$0.338^{+0.058}_{-0.200}$}\\
    & & & \\
   \hline
   \multirow{2}{*}{\textit{T}-dist. likelihood with $\nu$ = 13 (10\% larger tails)} & \multirow{2}{*}{\(0.589\pm 0.020\)} & \multirow{2}{*}{$0.804^{+0.140}_{-0.074}$} & \multirow{2}{*}{$0.336^{+0.057}_{-0.200}$}\\
    & & & \\
    \hline
\end{tabular}
\caption{These are the 68\% credible intervals for the \(\Lambda\)CDM model using CMB lensing measurements. The table compares the results for a Gaussian likelihood to \textit{t}-distribution likelihoods derived from a 1\% and 10\% increase of a Gaussian's 2$\sigma$ region.}
\label{table:LensingMargResults}
\end{table}

We also applied this sensitivity analysis to CMB lensing data for the \(\Lambda\)CDM model, which resulted in the parameter constraints outlined in Table~(\ref{table:LensingMargResults}). The CMB lensing results exhibited almost no variation in the parameter constraints with regards to the form of the likelihood, with only up to $0.02\sigma$ differences. We further confirmed that CMB lensing with the aggressive multipole range (with larger correction factors due to larger $n_d$) had results consistent with that found for the conservative range. 

\subsection{CMB \textit{T}\&\textit{P}}

 \begin{table}[h!]
\centering
\begin{tabular}{|c|c|c|c|c|c|c|c|c|}
    \hline
   \multicolumn{9}{|c|}{\textit{Planck} CMB \textit{TTTEEE} + \code{bflike} low$TEB$} \\
   \hline
    & \multicolumn{4}{c|}{Original \textit{Planck}, Gaussian likelihood} & \multicolumn{4}{c|}{\textit{T}-dist.~likelihood, 1\% larger tails}\\
    \hline 
     Parameter &  Mean & $\pm\sigma$ & $\pm2\sigma$ & $\pm3 \sigma$ & Mean & $\pm\sigma$ & $\pm2\sigma$ & $\pm3\sigma$ \\
    \hline
    \multirow{2}{*}{\(\Omega_\mathrm{b} h^2\)} & 
    \multirow{2}{*}{\(0.02243\)} &
    \multirow{2}{*}{\(0.00015\)} & 
    \multirow{2}{*}{\(0.00030\)} &
    \multirow{2}{*}{\(^{+0.00047}_{-0.00046}\)} & 
    \multirow{2}{*}{\(0.02242\)} &
    \multirow{2}{*}{\( 0.00016\)} & 
    \multirow{2}{*}{\(^{+0.00031}_{-0.00031}\)} &
    \multirow{2}{*}{\(^{+0.00047}_{-0.00046}\)} \\
    & & & & & & & & \\

    \multirow{2}{*}{\(\Omega_\mathrm{c} h^2\)} & 
    \multirow{2}{*}{\(0.1195\)} &
    \multirow{2}{*}{\(0.0014\)} & 
    \multirow{2}{*}{\(0.0028\)} &
    \multirow{2}{*}{\(^{+0.0045}_{-0.0042}\)} &
    \multirow{2}{*}{\(0.1195\)} &
    \multirow{2}{*}{\( 0.0014\)} & 
    \multirow{2}{*}{\(0.0028\)} &
    \multirow{2}{*}{\(^{+0.0042}_{-0.0043}\)} \\
    & & & & & & & & \\

    \multirow{2}{*}{\(100\theta_{\mathrm{MC}}\)} & 
    \multirow{2}{*}{\(1.04097\)} & 
    \multirow{2}{*}{\(0.00032\)} &
    \multirow{2}{*}{\(^{+0.00063}_{-0.00062}\)} &
    \multirow{2}{*}{\(^{+0.0010}_{-0.00093}\)} & 
    \multirow{2}{*}{\(1.04097\)} &
    \multirow{2}{*}{\( 0.00032\)} & 
    \multirow{2}{*}{\(^{+0.00062}_{-0.00063}\)} &
    \multirow{2}{*}{\(^{+0.00095}_{-0.00099}\)} \\
    & & & & & & & & \\

    \multirow{2}{*}{\(\tau\)} & 
    \multirow{2}{*}{\(0.0780\)} &
    \multirow{2}{*}{\(_{-0.0145}^{+0.0146}\)} & 
    \multirow{2}{*}{\(_{-0.0292}^{+0.0288}\)} &
    \multirow{2}{*}{\(_{-0.0458}^{+0.0436}\)} & 
    \multirow{2}{*}{\(0.0757\)} &
    \multirow{2}{*}{\(_{-0.0146}^{+0.0148}\)} & 
    \multirow{2}{*}{\(^{+0.0292}_{-0.0303}\)} &
    \multirow{2}{*}{\(^{+0.0436}_{-0.0502}\)} \\
    & & & & & & & & \\

    \multirow{2}{*}{ln(\(10^{10}A_\mathrm{s})\)} & 
    \multirow{2}{*}{\(3.090\)} & 
    \multirow{2}{*}{\(0.029\)} & 
    \multirow{2}{*}{\(^{+0.056}_{-0.057}\)} &
    \multirow{2}{*}{\(^{+0.084}_{-0.090}\)} & 
    \multirow{2}{*}{\(3.086\)} &
    \multirow{2}{*}{\( 0.029\)} & 
    \multirow{2}{*}{\(^{+0.056}_{-0.060}\)} &
    \multirow{2}{*}{\(^{+0.086}_{-0.097}\)} \\
    & & & & & & & & \\

    \multirow{2}{*}{\(n_\mathrm{s}\)} & 
    \multirow{2}{*}{\(0.9671\)} & 
    \multirow{2}{*}{\(0.0046\)} & 
    \multirow{2}{*}{\( 0.0090\)} &
    \multirow{2}{*}{\(^{+0.0138}_{-0.0137}\)} & 
    \multirow{2}{*}{\(0.9671\)} &
    \multirow{2}{*}{\( 0.0047\)} & 
    \multirow{2}{*}{\(^{+0.0093}_{-0.0090}\)} &
    \multirow{2}{*}{\(0.0139\)} \\
    & & & & & & & & \\
    \hline
    \multirow{2}{*}{\(H_0\)} & 
    \multirow{2}{*}{\(67.59\)} & 
    \multirow{2}{*}{\(0.63\)} & 
    \multirow{2}{*}{\(1.24\)} &
    \multirow{2}{*}{\(_{-1.91}^{+1.93}\)} & 
    \multirow{2}{*}{\(67.58\)} &
    \multirow{2}{*}{\( 0.65\)} & 
    \multirow{2}{*}{\(^{+1.27}_{-1.26}\)} &
    \multirow{2}{*}{\(^{+2.00}_{-1.86}\)} \\
    & & & & & & & & \\
     
     \multirow{2}{*}{\(\Omega_\Lambda\)} & 
     \multirow{2}{*}{\(0.6877\)} & 
     \multirow{2}{*}{\(0.0087\)} & 
     \multirow{2}{*}{\(_{-0.0175}^{+0.0166}\)} &
    \multirow{2}{*}{\(_{-0.0278}^{+0.0251}\)} & 
    \multirow{2}{*}{\( 0.6876\)} &
    \multirow{2}{*}{\( 0.0089\)} & 
    \multirow{2}{*}{\(^{+0.0169}_{-0.0176}\)} &
    \multirow{2}{*}{\(^{+0.0257}_{-0.0264}\)} \\
    & & & & & & & & \\
    
    \multirow{2}{*}{\(\Omega_\mathrm{m}\)} & 
    \multirow{2}{*}{\( 0.312\)} & 
    \multirow{2}{*}{\(0.009\)} & 
    \multirow{2}{*}{\(^{+0.018}_{-0.017}\)} &
    \multirow{2}{*}{\(^{+0.028}_{-0.025}\)} & 
    \multirow{2}{*}{\(0.312\)} &
    \multirow{2}{*}{\( 0.009\)} & 
    \multirow{2}{*}{\(^{+0.018}_{-0.017}\)} &
    \multirow{2}{*}{\(^{+0.026}_{-0.026}\)} \\
   & & & & & & & & \\
     
     \multirow{2}{*}{\(\sigma_8\)} & 
     \multirow{2}{*}{\(0.8290\)} & 
     \multirow{2}{*}{\(_{-0.0115}^{+0.0116}\)} & 
     \multirow{2}{*}{\(_{-0.0232}^{+0.0230}\)} &
    \multirow{2}{*}{\(_{-0.0352}^{+0.0348}\)} &
    \multirow{2}{*}{\(0.8271\)} &
    \multirow{2}{*}{\(_{-0.0117}^{+0.0118}\)} & 
    \multirow{2}{*}{\(^{+0.0228}_{-0.0237}\)} &
    \multirow{2}{*}{\(^{+0.0353}_{-0.0374}\)} \\
    & & & & & & & & \\
     
     \multirow{2}{*}{\(S_8\)} & 
     \multirow{2}{*}{\(0.846\)} & 
     \multirow{2}{*}{\(0.017\)} & 
     \multirow{2}{*}{\({0.034}\)} &
    \multirow{2}{*}{\(^{+0.053}_{-0.050}\)} &  
    \multirow{2}{*}{\(0.844\)} &
    \multirow{2}{*}{\( 0.018\)} & 
    \multirow{2}{*}{\(^{+0.034}_{-0.035}\)} &
    \multirow{2}{*}{\(^{+0.052}_{-0.049}\)} \\
    & & & & & & & & \\
     
     \multirow{2}{*}{\(\sigma_8\Omega_\mathrm{m}^{0.25}\)} & 
     \multirow{2}{*}{\(0.6197\)} & 
     \multirow{2}{*}{\(_{-0.0100}^{+0.0098}\)} & 
     \multirow{2}{*}{\(_{-0.0197}^{+0.0196}\)} &
    \multirow{2}{*}{\(_{-0.0291}^{+0.0300}\)} & 
    \multirow{2}{*}{\(0.6184\)} &
    \multirow{2}{*}{\(_{-0.0101}^{+0.0100}\)} & 
    \multirow{2}{*}{\(^{+0.0195}_{-0.0202}\)} &
    \multirow{2}{*}{\(0.0294\)} \\
    & & & & & & & & \\
    
    \multirow{2}{*}{\(z_\mathrm{re}\)} & 
    \multirow{2}{*}{\(9.88\)} & 
     \multirow{2}{*}{\(_{-1.22}^{+1.38}\)} & 
     \multirow{2}{*}{\(_{-2.74}^{+2.47}\)} &
    \multirow{2}{*}{\(_{-4.64}^{+3.61}\)} &  
    \multirow{2}{*}{\(9.67\)} &
    \multirow{2}{*}{\(^{+1.45}_{-1.19}\)} & 
    \multirow{2}{*}{\(^{+2.65}_{-2.73}\)} &
    \multirow{2}{*}{\(^{+3.64}_{-5.23}\)} \\
    & & & & & & & & \\
    
    \multirow{2}{*}{\(10^9A_\mathrm{s}e^{-2\tau}\)} & 
    \multirow{2}{*}{\(1.881\)} & 
     \multirow{2}{*}{\(0.012\)} & 
     \multirow{2}{*}{\(0.023\)} &
    \multirow{2}{*}{\(0.035\)} &   
    \multirow{2}{*}{\(1.881\)} &
    \multirow{2}{*}{\( 0.012\)} & 
    \multirow{2}{*}{\(0.023\)} &
    \multirow{2}{*}{\(^{+0.036}_{-0.034}\)} \\
    & & & & & & & & \\
    \hline 
\end{tabular}
\caption{Difference between the \(\Lambda\)CDM model parameter constraints when using \textit{Planck} CMB \code{Plik} high-$l$ \textit{TT},\textit{TE},\textit{EE} and \code{bflike} low-$l$ \textit{TT},\textit{TE},\textit{EE},\textit{BB} measurements for a Gaussian likelihood versus a \textit{t}-distribution likelihood with 1\% larger tails ($\nu=104$). Included are the 68\%, 95\%, and 99.7\% credible intervals. We used the units of km $\mathrm{s}^{-1}\mathrm{Mpc}^{-1}$ for $H_0$ and defined \(S_8\equiv \sigma_8(\Omega_\mathrm{m}/0.3)^{0.5}\). In the first grouping of rows are the base parameters used in our MCMC analysis and the bottom group are the derived parameters.
}
\label{table:TP1margtdistSigmaResults}
\end{table}

\begin{table}[h!]
\centering
\begin{tabular}{|c|c|c|c|c|c|c|c|c|}
    \hline
   \multicolumn{9}{|c|}{\textit{Planck} CMB $TTTEEE$ + \code{bflike} low$TEB$} \\
   \hline
    & \multicolumn{4}{c|}{Original \textit{Planck}, Gaussian likelihood} & \multicolumn{4}{c|}{\textit{T}-dist.~likelihood, 10\% larger tails}  \\
    \hline 
     Parameter &  Mean & $\pm\sigma$ & $\pm2\sigma$ & $\pm3\sigma$ & Mean & $\pm\sigma$ & $\pm2\sigma$ & $\pm3\sigma$ \\
    \hline
    \multirow{2}{*}{\(\Omega_\mathrm{b} h^2\)} & 
    \multirow{2}{*}{\(0.02243\)} &
    \multirow{2}{*}{\(0.00015\)} & 
    \multirow{2}{*}{\(0.00030\)} &
    \multirow{2}{*}{\(^{+0.00047}_{-0.00046}\)} & 
    \multirow{2}{*}{\(0.02242\)} &
    \multirow{2}{*}{\( 0.00016\)} & 
    \multirow{2}{*}{\(^{+0.00031}_{-0.00030}\)} &
    \multirow{2}{*}{\(^{+0.00047}_{-0.00048}\)} \\
    & & & & & & & & \\

    \multirow{2}{*}{\(\Omega_\mathrm{c} h^2\)} & 
    \multirow{2}{*}{\(0.1195\)} &
    \multirow{2}{*}{\(0.0014\)} & 
    \multirow{2}{*}{\(0.0028\)} &
    \multirow{2}{*}{\(^{+0.0045}_{-0.0042}\)} &
    \multirow{2}{*}{\(0.1195\)} &
    \multirow{2}{*}{\( 0.0014\)} & 
    \multirow{2}{*}{\(0.0028\)} &
    \multirow{2}{*}{\(0.0042\)} \\
    & & & & & & & & \\

    \multirow{2}{*}{\(100\theta_{\mathrm{MC}}\)} & 
    \multirow{2}{*}{\(1.04097\)} & 
    \multirow{2}{*}{\(0.00032\)} &
    \multirow{2}{*}{\(^{+0.00063}_{-0.00062}\)} &
    \multirow{2}{*}{\(^{+0.0010}_{-0.00093}\)} & 
    \multirow{2}{*}{\(1.04097\)} &
    \multirow{2}{*}{\( 0.00032\)} & 
    \multirow{2}{*}{\(0.00063\)} &
    \multirow{2}{*}{\(^{+0.00093}_{-0.00097}\)} \\
    & & & & & & & & \\

    \multirow{2}{*}{\(\tau\)} & 
    \multirow{2}{*}{\(0.0780\)} &
    \multirow{2}{*}{\(_{-0.0145}^{+0.0146}\)} & 
    \multirow{2}{*}{\(_{-0.0292}^{+0.0288}\)} &
    \multirow{2}{*}{\(_{-0.0458}^{+0.0436}\)} & 
    \multirow{2}{*}{\(0.0759\)} &
    \multirow{2}{*}{\(_{-0.0145}^{+0.0147}\)} & 
    \multirow{2}{*}{\(0.0290\)} &
    \multirow{2}{*}{\(^{+0.0434}_{-0.0451}\)} \\
    & & & & & & & & \\

    \multirow{2}{*}{ln(\(10^{10}A_\mathrm{s})\)} & 
    \multirow{2}{*}{\(3.090\)} & 
    \multirow{2}{*}{\(0.029\)} & 
    \multirow{2}{*}{\(^{+0.056}_{-0.057}\)} &
    \multirow{2}{*}{\(^{+0.084}_{-0.090}\)} & 
    \multirow{2}{*}{\(3.086\)} &
    \multirow{2}{*}{\( 0.029\)} & 
    \multirow{2}{*}{\(^{+0.056}_{-0.057}\)} &
    \multirow{2}{*}{\(^{+0.086}_{-0.089}\)} \\
    & & & & & & & & \\

    \multirow{2}{*}{\(n_\mathrm{s}\)} & 
    \multirow{2}{*}{\(0.9671\)} & 
    \multirow{2}{*}{\(0.0046\)} & 
    \multirow{2}{*}{\( 0.0090\)} &
    \multirow{2}{*}{\(^{+0.0138}_{-0.0137}\)} & 
    \multirow{2}{*}{\(0.9672\)} &
    \multirow{2}{*}{\( 0.0046\)} & 
    \multirow{2}{*}{\(^{+0.0091}_{-0.0091}\)} &
    \multirow{2}{*}{\(^{+0.0137}_{-0.0136}\)} \\
    & & & & & & & & \\
    \hline
    \multirow{2}{*}{\(H_0\)} & 
    \multirow{2}{*}{\(67.59\)} & 
    \multirow{2}{*}{\(0.63\)} & 
    \multirow{2}{*}{\(1.24\)} &
    \multirow{2}{*}{\(_{-1.91}^{+1.93}\)} & 
    \multirow{2}{*}{\(67.60\)} &
    \multirow{2}{*}{\( 0.64\)} & 
    \multirow{2}{*}{\(^{+1.25}_{-1.24}\)} &
    \multirow{2}{*}{\(^{+1.95}_{-1.89}\)} \\
    & & & & & & & & \\
     
     \multirow{2}{*}{\(\Omega_\Lambda\)} & 
     \multirow{2}{*}{\(0.6877\)} & 
     \multirow{2}{*}{\(0.0087\)} & 
     \multirow{2}{*}{\(_{-0.0175}^{+0.0166}\)} &
    \multirow{2}{*}{\(_{-0.0278}^{+0.0251}\)} & 
    \multirow{2}{*}{\(0.6879\)} &
    \multirow{2}{*}{\( 0.0087\)} & 
    \multirow{2}{*}{\(^{+0.0167}_{-0.0174}\)} &
    \multirow{2}{*}{\(^{+0.0255}_{-0.0269}\)} \\
    & & & & & & & & \\
    
    \multirow{2}{*}{\(\Omega_\mathrm{m}\)} & 
    \multirow{2}{*}{\( 0.312\)} & 
    \multirow{2}{*}{\(0.009\)} & 
    \multirow{2}{*}{\(^{+0.018}_{-0.017}\)} &
    \multirow{2}{*}{\(^{+0.028}_{-0.025}\)} & 
    \multirow{2}{*}{\(0.312\)} &
    \multirow{2}{*}{\( 0.009\)} & 
    \multirow{2}{*}{\(0.017\)} &
    \multirow{2}{*}{\(^{+0.027}_{-0.025}\)} \\
   & & & & & & & & \\
     
     \multirow{2}{*}{\(\sigma_8\)} & 
     \multirow{2}{*}{\(0.8290\)} & 
     \multirow{2}{*}{\(_{-0.0115}^{+0.0116}\)} & 
     \multirow{2}{*}{\(_{-0.0232}^{+0.0230}\)} &
    \multirow{2}{*}{\(_{-0.0352}^{+0.0348}\)} &
    \multirow{2}{*}{\(0.8271\)} &
    \multirow{2}{*}{\(_{-0.0115}^{+0.0117}\)} & 
    \multirow{2}{*}{\(0.0227\)} &
    \multirow{2}{*}{\(^{+0.0343}_{-0.0355}\)} \\
    & & & & & & & & \\
     
     \multirow{2}{*}{\(S_8\)} & 
     \multirow{2}{*}{\(0.846\)} & 
     \multirow{2}{*}{\(0.017\)} & 
     \multirow{2}{*}{\({0.034}\)} &
    \multirow{2}{*}{\(^{+0.053}_{-0.050}\)} &  
    \multirow{2}{*}{\( 0.844\)} &
    \multirow{2}{*}{\( 0.017\)} & 
    \multirow{2}{*}{\(^{+0.034}_{-0.033}\)} &
    \multirow{2}{*}{\(^{+0.051}_{-0.050}\)} \\
    & & & & & & & & \\
     
     \multirow{2}{*}{\(\sigma_8\Omega_\mathrm{m}^{0.25}\)} & 
     \multirow{2}{*}{\(0.6197\)} & 
     \multirow{2}{*}{\(_{-0.0100}^{+0.0098}\)} & 
     \multirow{2}{*}{\(_{-0.0197}^{+0.0196}\)} &
    \multirow{2}{*}{\(_{-0.0291}^{+0.0300}\)} & 
    \multirow{2}{*}{\(0.6182\)} &
    \multirow{2}{*}{\( 0.0099\)} & 
    \multirow{2}{*}{\(^{+0.0193}_{-0.0195}\)} &
    \multirow{2}{*}{\(^{+0.0289}_{-0.0294}\)} \\
    & & & & & & & & \\
    
    \multirow{2}{*}{\(z_\mathrm{re}\)} & 
    \multirow{2}{*}{\(9.88\)} & 
     \multirow{2}{*}{\(_{-1.22}^{+1.38}\)} & 
     \multirow{2}{*}{\(_{-2.74}^{+2.47}\)} &
    \multirow{2}{*}{\(_{-4.64}^{+3.61}\)} &  
    \multirow{2}{*}{\(9.69\)} &
    \multirow{2}{*}{\(^{+1.40}_{-1.23}\)} & 
    \multirow{2}{*}{\(^{+2.60}_{-2.64}\)} &
    \multirow{2}{*}{\(^{+3.58}_{-4.68}\)} \\
    & & & & & & & & \\
    
    \multirow{2}{*}{\(10^9A_\mathrm{s}e^{-2\tau}\)} & 
    \multirow{2}{*}{\(1.881\)} & 
     \multirow{2}{*}{\(0.012\)} & 
     \multirow{2}{*}{\(0.023\)} &
    \multirow{2}{*}{\(0.035\)} &   
    \multirow{2}{*}{\(1.881\)} &
    \multirow{2}{*}{\( 0.012\)} & 
    \multirow{2}{*}{\(0.023\)} &
    \multirow{2}{*}{\(^{+0.035}_{-0.037}\)} \\
    & & & & & & & & \\
    \hline 
\end{tabular}
\caption{Comparison of parameter constraints for the \(\Lambda\)CDM model using \textit{Planck} CMB temperature and polarization measurements for a Gaussian likelihood and a \textit{t}-distribution likelihood with 10\% more probability in the tails ($\nu=13$). The uncertainties are the 68\%, 95\%, and 99.7\% intervals, the units of \(H_0\) are km $\mathrm{s}^{-1}\mathrm{Mpc}^{-1}$, and \(S_8\equiv \sigma_8(\Omega_\mathrm{m}/0.3)^{0.5}\). In the first section of rows are the base parameters sampled over in MCMC, and the bottom section are the derived parameters.}
\label{table:TP10margtdistSigmaResults}
\end{table}

\begin{table}[h!]
\centering
\begin{tabular}{|c|c|c|c|c|}
    \hline
   \multicolumn{5}{|c|}{\textit{Planck} CMB low$TEB$} \\
   \hline
    \multirow{3}{*}{Parameter} & Original \textit{Planck} & \textit{T}-dist. likelihood  & \textit{T}-dist. likelihood & \textit{T}-dist. likelihood \\
     & method with &  with $\nu$ = 213  &  with $\nu$ = 104 &  with $\nu$ = 13 \\
     & Gaussian likelihood & (0.3\% larger tails) & (1\% larger tails) & (10\% larger tails) \\
    \hline

    \multirow{2}{*}{\(\tau\)} &
    \multirow{2}{*}{\(0.062^{+0.022}_{-0.017}\)} &
    \multirow{2}{*}{\(0.060^{+0.022}_{-0.018}\)} &
    \multirow{2}{*}{\(0.059^{+0.022}_{-0.018}\)} &
    \multirow{2}{*}{\(0.059^{+0.022}_{-0.017}\)} \\
    & & & & \\

    \multirow{2}{*}{ln(\(10^{10}A_\mathrm{s})\)} & 
    \multirow{2}{*}{\(2.966\pm 0.056\)} &
    \multirow{2}{*}{\(2.927\pm 0.061\)} &
    \multirow{2}{*}{\(2.925\pm 0.060\)} &
    \multirow{2}{*}{\(2.925\pm 0.060\)} \\
    & & & & \\

    \multirow{2}{*}{\(10^9A_\mathrm{s}e^{-2\tau}\)} & 
    \multirow{2}{*}{\(1.718^{+0.078}_{-0.091}\)} &
    \multirow{2}{*}{\(1.660\pm 0.090\)} &
    \multirow{2}{*}{\(1.657^{+0.083}_{-0.096}\)} &
    \multirow{2}{*}{\(1.658^{+0.082}_{-0.097}\)} \\
    & & & & \\
    \hline 
\end{tabular}
\caption{For \textit{Planck} CMB low-$l$ $TT$,$TE$,$EE$,$BB$ measurements, the \(\Lambda\)CDM model parameter constraints are compared for a Gaussian likelihood and\textit{ t}-distribution likelihoods that have 0.3\%, 1\% and 10\% larger 2$\sigma$ regions than the Gaussian. These are the 68\% credible intervals. }
\label{table:LFIonlymargtdist}
\end{table}
 
Fitting to the CMB temperature and polarization data, we again replaced the Gaussian likelihood with a multivariate \textit{t}-distribution for different choices of degrees of freedom, $\nu$. The results for $TTTEEE$ and \code{bflike} low$TEB$ are shown in Tables~(\ref{table:TP1margtdistSigmaResults}) and (\ref{table:TP10margtdistSigmaResults}), then for only low$TEB$ in Table~(\ref{table:LFIonlymargtdist}). We found for $TTTEEE$ + low$TEB$ that parameters related to optical depth and the overall power spectrum amplitude $\{\tau, \ln(10^{10}A_\mathrm{s}), \sigma_8, S_8, \sigma_8\Omega_\mathrm{m}^{0.25}, z_\mathrm{re}, 10^9A_\mathrm{s}e^{-2\tau}\}$ exhibited similar behaviour, ranging between a $0.1$--$0.2\sigma$ difference from the \textit{t}-distributions to the Gaussian. This corresponds to a lower value of $\sigma_8$ and $S_8$, which slightly reduces the $S_8$ tension. The other parameters did not differ significantly from the original Gaussian likelihood, with a maximum discrepancy of 0.03$\sigma$. For the low$TEB$ \textit{t}-distribution results, we found that $\tau$ changed by $0.15\sigma$ whereas ln(\(10^{10}A_\mathrm{s})\) and $10^9A_\mathrm{s}e^{-2\tau}$ had a $0.7\sigma$ difference from the Gaussian case. By increasing power in the tails of the distribution, we obtain a lower estimate of $\tau$ and $A_\mathrm{s}$. This prompted us to investigate another choice of degrees of freedom, $\nu=213$, to confirm this trend in parameter constraints. This corresponds to a 0.3\% increase of the Gaussian's $2\sigma$ region, so it is a mid-point between the Gaussian case and the 1\% increase. Indeed, it does show the same behaviour of decreased mean values of $\tau$ and $A_\mathrm{s}$, suggesting that the change in constraints may be a discontinuous one caused by changing the likelihood from Gaussian to a $t$-distribution. We should note that the polarization of the CMB at large angular scales is dominated by noise and, because there is a strong degeneracy between $\tau$ and $A_\mathrm{s}$, these parameters are sensitive to this noisy signal. Since we use the variance of the map to infer cosmological parameters, any residual or additional noise not captured by the noise covariance matrix will be interpreted as additional signal, thus causing a systematic preference for larger values of $\tau$ and $A_\mathrm{s}$. Therefore, our results show that more weight in the tails of the distribution means more weight is given to the noise, lowering $10^9A_\mathrm{s}e^{-2\tau}$.

There is a well-known tendency for $\tau$ to shift between different CMB analyses, due to the difficulty of modelling the extremely large angular scales that are sensitive to $\tau$.
For instance, the official \textit{Planck} 2018 low-$l$ LFI (\code{bflike}) likelihood value is $\tau = 0.063\pm0.020$ \cite{Planck2018-05} and the combined \code{Commander} and \code{SimAll} result is $\tau=0.0506\pm0.0086$.
Natale et al. \cite{Natale:2020} reported $\tau = 0.069_{-0.012}^{+0.011}$ for their \textit{WMAP} + LFI likelihood, which roughly corresponds to a $0.9\sigma$ difference from the \textit{t}-distributions. Moreover, the \code{BeyondPlanck} LFI results are $\tau = 0.065\pm0.012$, which corresponds to a $0.5\sigma$ offset \cite{Paradiso2022}.
These shifts in $\tau$ are comparable to or somewhat larger than the shifts we find of $\Delta \tau = 0.003$, when switching from a Gaussian to a $t$-distribution.
 This also shows that $\tau$ is very sensitive to choices made in the likelihood, possibly explaining discrepancies seen in the literature.
However, interestingly, the overall amplitude $A_\mathrm{s} e^{-2\tau}$ and the matter fluctuation amplitude $A_\mathrm{s}$ are more sensitive to the form of the likelihood than $\tau$ for low-$l$ LFI data. When high-$\ell$ and HFI data are included this sensitivity decreases significantly.

\section{Conclusions}\label{conclusion}

We have examined how cosmological constraints from BAO and CMB measurements rely on the form of the likelihood used in the parameter inferences. Initially we considered how, when the covariance matrix is estimated through simulations, we can marginalize over the true covariance matrix instead of assuming a Gaussian likelihood with a scaled covariance. Then, we tested the effect of different choices of priors on the covariance matrix, including the Independence-Jeffreys prior and the frequentist matching prior. We found minimal differences between the different approaches when applied to current BAO and CMB lensing data from SDSS and Planck, respectively (less than $0.05\sigma$ differences). We further confirmed that the insensitivity to the form of the likelihood held regardless of whether we fit to the correlation function data or the compressed AP parameters. This demonstrates that the previously published constraints are robust to the change in the likelihood due to uncertainties in the covariance matrix, but future data analyses should consider using the $t$-distribution likelihood when the covariance is estimated from simulations, as it is more statistically robust. 

Second, we have performed a sensitivity test on the assumption of Gaussianity in the likelihood by examining how parameter constraints are affected by increasing the probability in the tails for a wider range of data. When constraining extensions to flat $\Lambda$CDM with BAO only, we find some sensitivity to using heavier tails with a multivariate $t$-distribution likelihood. In particular, $\Omega_\Lambda$ changed by $\sim 0.2\sigma$ for a \textit{t}-distribution with 1\% larger tails and $0.4\sigma$ for 10\% larger tails. To find this sensitivity, we must directly fit the correlation functions to obtain the parameter constraints rather than modifying the cosmological likelihood for the compressed AP parameters. This is because the compressed AP parameters assume a Gaussian correlation function likelihood~\cite{RossBOSS}. Additionally, we emphasize that we see these changes when considering open $\Lambda$CDM, which has some very weakly constrained parameters, which may be more susceptible to likelihood non-Gaussianity. This demonstrates how assumptions about the likelihood can influence parameter constraints, which may be more relevant in cases where the model is not well constrained by data.

When applying the sensitivity test to the CMB in the $\Lambda$CDM model, we found that the cosmological constraints were mostly insensitive to increased probability in the tails of the likelihood. However, for CMB temperature and polarization data, the $TTTEEE$ + \code{bflike} low$TEB$ results had $\{\tau$, ln(\(10^{10}A_\mathrm{s})\), $\sigma_8$, $S_8$, $\sigma_8\Omega_\mathrm{m}^{0.25}$, $z_\mathrm{re}$, $10^9A_\mathrm{s}e^{-2\tau}\}$ change by about $\sim0.1$--$0.2\sigma$ from the original Gaussian results. Moreover, the low$TEB$ (low-$l$ LFI) $\tau$ constraints differed by $0.15\sigma$ and both ln(\(10^{10}A_\mathrm{s})\) and $10^9A_\mathrm{s}e^{-2\tau}$ differed by 0.7$\sigma$. Given the long history of $\tau$ constraints being the most heavily influenced by systematic uncertainties, this parameter provides insights into possible issues in the data analysis. If we consider runs where only  the low$TEB$ likelihood is included, the constraint for a 1\% widening of the Gaussian's 2$\sigma$ region is \(\tau = 0.059^{+0.022}_{-0.018}\) and for a 10\% widening it is \(\tau = 0.059^{+0.022}_{-0.017}\). This is lower than the official \textit{Planck} 2018 result of $\tau = 0.063\pm0.020$ from using the low-$l$ LFI likelihood~\cite{Planck2018-05} and even lower than the \code{BeyondPlanck} LFI $\tau = 0.065\pm0.012$ \cite{Paradiso2022} and \textit{WMAP}+LFI $\tau = 0.069_{-0.012}^{+0.011}$~\cite{Natale:2020}. Therefore, if there are any unaccounted systematic effects that result in larger tails of the likelihood, $\tau$ and $A_\mathrm{s}$ would be shifted to a lower mean value than currently estimated. However, note that when high-$\ell$ and HFI data are included, the improved constraining power and decreased sensitivity to residuals help to break the $\tau$--$A_\mathrm{s}$ degeneracy. 
As a result, we only find small ($\sim0.15\sigma$) shifts in both of these parameters when using the \textit{t}-distribution on the high-$\ell$ and HFI data.

\acknowledgments

We would like to thank Julien Carron and Ashley Ross for their assistance in understanding the publicly available CMB lensing and BOSS post-reconstruction BAO data products, respectively.

This research was undertaken as part of the Perimeter Scholars International (PSI) Master's program at the Perimeter Institute for Theoretical Physics. Research at Perimeter Institute is supported in part by the Government of Canada through the Department of Innovation, Science and Economic Development Canada and by the Province of Ontario through the Ministry of Colleges and Universities.

This research was enabled in part by support provided by Compute Ontario (computeontario.ca) and the Digital Research Alliance of Canada (alliancecan.ca).

SP acknowledges support from the Canadian Government through a New Frontiers in Research Fund (NFRF) Exploration grant.
WP acknowledges the support of the Natural Sciences and Engineering Research Council of Canada (NSERC), [funding reference number RGPIN-2019-03908] and from the Canadian Space Agency.
JK was supported by NSERC through the Vanier Canada Graduate Scholarship. 
AK was supported as a CITA National Fellow by the Natural Sciences and Engineering Research Council of Canada (NSERC), funding reference \#DIS-2022-568580.









\addcontentsline{toc}{section}{References}
\bibliographystyle{JHEP}
\bibliography{References}

\providecommand{\href}[2]{#2}\begingroup\raggedright\begin{thebibliography}{10}

\bibitem{BOSS}
K.S.~{Dawson}, D.J.~{Schlegel}, C.P.~{Ahn}, S.F.~{Anderson}, {\'E}.~{Aubourg},
  S.~{Bailey} et~al., \emph{{The Baryon Oscillation Spectroscopic Survey of
  SDSS-III}}, \href{https://doi.org/10.1088/0004-6256/145/1/10}{\emph{"The
  Astronomical Journal"} {\bfseries 145} (2013) 10}
  [\href{https://arxiv.org/abs/1208.0022}{{\ttfamily 1208.0022}}].

\bibitem{eBOSS}
K.S.~{Dawson}, J.-P.~{Kneib}, W.J.~{Percival}, S.~{Alam}, F.D.~{Albareti},
  S.F.~{Anderson} et~al., \emph{{The SDSS-IV Extended Baryon Oscillation
  Spectroscopic Survey: Overview and Early Data}},
  \href{https://doi.org/10.3847/0004-6256/151/2/44}{\emph{"Astron. J."}
  {\bfseries 151} (2016) 44}
  [\href{https://arxiv.org/abs/1508.04473}{{\ttfamily 1508.04473}}].

\bibitem{DES2021CosmoConstraints}
{\scshape DES} collaboration, \emph{{Dark Energy Survey Year 3 results:
  Cosmological constraints from galaxy clustering and weak lensing}},
  \href{https://doi.org/10.1103/PhysRevD.105.023520}{\emph{Phys. Rev. D}
  {\bfseries 105} (2022) 023520}
  [\href{https://arxiv.org/abs/2105.13549}{{\ttfamily 2105.13549}}].

\bibitem{Planck2018cosmoparams}
{\scshape Planck} collaboration, \emph{{Planck 2018 results. VI. Cosmological
  parameters}},
  \href{https://doi.org/10.1051/0004-6361/201833910}{\emph{Astron. Astrophys.}
  {\bfseries 641} (2020) A6}
  [\href{https://arxiv.org/abs/1807.06209}{{\ttfamily 1807.06209}}].

\bibitem{Pantheon2022}
D.~Brout et~al., \emph{{The Pantheon+ Analysis: Cosmological Constraints}},
  \href{https://doi.org/10.3847/1538-4357/ac8e04}{\emph{Astrophys. J.}
  {\bfseries 938} (2022) 110}
  [\href{https://arxiv.org/abs/2202.04077}{{\ttfamily 2202.04077}}].

\bibitem{Union2023}
D.~{Rubin}, G.~{Aldering}, M.~{Betoule}, A.~{Fruchter}, X.~{Huang}, A.G.~{Kim}
  et~al., \emph{{Union Through UNITY: Cosmology with 2,000 SNe Using a Unified
  Bayesian Framework}},
  \href{https://doi.org/10.48550/arXiv.2311.12098}{\emph{arXiv e-prints} (2023)
  arXiv:2311.12098} [\href{https://arxiv.org/abs/2311.12098}{{\ttfamily
  2311.12098}}].

\bibitem{Hubbletension}
A.G.~Riess, \emph{{The Expansion of the Universe is Faster than Expected}},
  \href{https://doi.org/10.1038/s42254-019-0137-0}{\emph{Nature Rev. Phys.}
  {\bfseries 2} (2019) 10} [\href{https://arxiv.org/abs/2001.03624}{{\ttfamily
  2001.03624}}].

\bibitem{SDSSeBOSS}
S.~Alam, M.~Aubert, S.~Avila, C.~Balland, J.E.~Bautista et~al., \emph{Completed
  {SDSS-IV} extended {B}aryon {O}scillation {S}pectroscopic {S}urvey:
  Cosmological implications from two decades of spectroscopic surveys at the
  {A}pache {P}oint {O}bservatory},
  \href{https://doi.org/10.1103/PhysRevD.103.083533}{\emph{Phys. Rev. D}
  {\bfseries 103} (2021) 083533}.

\bibitem{KiDS2021}
C.~Heymans et~al., \emph{{KiDS-1000 Cosmology: Multi-probe weak gravitational
  lensing and spectroscopic galaxy clustering constraints}},
  \href{https://doi.org/10.1051/0004-6361/202039063}{\emph{Astron. Astrophys.}
  {\bfseries 646} (2021) A140}
  [\href{https://arxiv.org/abs/2007.15632}{{\ttfamily 2007.15632}}].

\bibitem{HSC:Li2023}
X.~Li et~al., \emph{{Hyper Suprime-Cam Year 3 results: Cosmology from cosmic
  shear two-point correlation functions}},
  \href{https://doi.org/10.1103/PhysRevD.108.123518}{\emph{Phys. Rev. D}
  {\bfseries 108} (2023) 123518}
  [\href{https://arxiv.org/abs/2304.00702}{{\ttfamily 2304.00702}}].

\bibitem{KiDSandDES:2023}
{Dark Energy Survey and Kilo-Degree Survey Collaboration}, T.M.C.~{Abbott},
  M.~{Aguena}, A.~{Alarcon}, O.~{Alves}, A.~{Amon} et~al., \emph{{DES Y3 +
  KiDS-1000: Consistent cosmology combining cosmic shear surveys}},
  \href{https://doi.org/10.21105/astro.2305.17173}{\emph{The Open Journal of
  Astrophysics} {\bfseries 6} (2023) 36}
  [\href{https://arxiv.org/abs/2305.17173}{{\ttfamily 2305.17173}}].

\bibitem{desi2019}
M.~{Levi}, L.E.~{Allen}, A.~{Raichoor}, C.~{Baltay}, S.~{BenZvi}, F.~{Beutler}
  et~al., \emph{{The Dark Energy Spectroscopic Instrument (DESI)}},  in
  \emph{Bulletin of the American Astronomical Society}, vol.~51, p.~57, Sept.,
  2019, \href{https://doi.org/10.48550/arXiv.1907.10688}{DOI}
  [\href{https://arxiv.org/abs/1907.10688}{{\ttfamily 1907.10688}}].

\bibitem{euclid}
R.~{Laureijs}, J.~{Amiaux}, S.~{Arduini}, J.L.~{Augu{\`e}res}, J.~{Brinchmann},
  R.~{Cole} et~al., \emph{{Euclid Definition Study Report}},
  \href{https://doi.org/10.48550/arXiv.1110.3193}{\emph{arXiv e-prints} (2011)
  arXiv:1110.3193} [\href{https://arxiv.org/abs/1110.3193}{{\ttfamily
  1110.3193}}].

\bibitem{WMAP:2003}
{\scshape WMAP} collaboration, \emph{{Wilkinson Microwave Anisotropy Probe
  (WMAP) first year observations: TE polarization}},
  \href{https://doi.org/10.1086/377219}{\emph{Astrophys. J. Suppl.} {\bfseries
  148} (2003) 161} [\href{https://arxiv.org/abs/astro-ph/0302213}{{\ttfamily
  astro-ph/0302213}}].

\bibitem{WMAP:2013}
G.~{Hinshaw}, D.~{Larson}, E.~{Komatsu}, D.N.~{Spergel}, C.L.~{Bennett},
  J.~{Dunkley} et~al., \emph{{Nine-year Wilkinson Microwave Anisotropy Probe
  (WMAP) Observations: Cosmological Parameter Results}},
  \href{https://doi.org/10.1088/0067-0049/208/2/19}{\emph{Astrophysical Journal
  Supplement Series} {\bfseries 208} (2013) 19}
  [\href{https://arxiv.org/abs/1212.5226}{{\ttfamily 1212.5226}}].

\bibitem{Planck:2013pxb}
{\scshape Planck} collaboration, \emph{{Planck 2013 results. XVI. Cosmological
  parameters}},
  \href{https://doi.org/10.1051/0004-6361/201321591}{\emph{Astron. Astrophys.}
  {\bfseries 571} (2014) A16}
  [\href{https://arxiv.org/abs/1303.5076}{{\ttfamily 1303.5076}}].

\bibitem{Tristram:2020wbi}
M.~Tristram et~al., \emph{{Planck constraints on the tensor-to-scalar ratio}},
  \href{https://doi.org/10.1051/0004-6361/202039585}{\emph{Astron. Astrophys.}
  {\bfseries 647} (2021) A128}
  [\href{https://arxiv.org/abs/2010.01139}{{\ttfamily 2010.01139}}].

\bibitem{Paradiso2022}
S.~Paradiso et~al., \emph{{BeyondPlanck XII. Cosmological parameter constraints
  with end-to-end error propagation}},
  \href{https://arxiv.org/abs/2205.10104}{{\ttfamily 2205.10104}}.

\bibitem{Cobaya}
J.~Torrado and A.~Lewis, \emph{{Cobaya: Code for Bayesian Analysis of
  hierarchical physical models}},
  \href{https://doi.org/10.1088/1475-7516/2021/05/057}{\emph{JCAP} {\bfseries
  05} (2021) 057} [\href{https://arxiv.org/abs/2005.05290}{{\ttfamily
  2005.05290}}].

\bibitem{percivalmatching}
W.J.~Percival, O.~Friedrich, E.~Sellentin and A.~Heavens, \emph{{Matching
  {B}ayesian and frequentist coverage probabilities when using an approximate
  data covariance matrix}},
  \href{https://doi.org/10.1093/mnras/stab3540}{\emph{Monthly Notices of the
  Royal Astronomical Society} {\bfseries 510} (2022) 3207}
  [\href{https://arxiv.org/abs/2108.10402}{{\ttfamily 2108.10402}}].

\bibitem{Percival2006}
W.J.~{Percival} and M.L.~{Brown}, \emph{{Likelihood techniques for the combined
  analysis of CMB temperature and polarization power spectra}},
  \href{https://doi.org/10.1111/j.1365-2966.2006.10910.x}{\emph{"Monthly
  Notices of the Royal Astronomical Society"} {\bfseries 372} (2006) 1104}
  [\href{https://arxiv.org/abs/astro-ph/0604547}{{\ttfamily
  astro-ph/0604547}}].

\bibitem{HamimecheLewis2008}
S.~Hamimeche and A.~Lewis, \emph{Likelihood analysis of cmb temperature and
  polarization power spectra},
  \href{https://doi.org/10.1103/PhysRevD.77.103013}{\emph{Phys. Rev. D}
  {\bfseries 77} (2008) 103013}.

\bibitem{DodelsonSchneider2013}
S.~{Dodelson} and M.D.~{Schneider}, \emph{{The effect of covariance estimator
  error on cosmological parameter constraints}},
  \href{https://doi.org/10.1103/PhysRevD.88.063537}{\emph{"Phys. Rev. D"}
  {\bfseries 88} (2013) 063537}
  [\href{https://arxiv.org/abs/1304.2593}{{\ttfamily 1304.2593}}].

\bibitem{Taylor2013}
A.~{Taylor}, B.~{Joachimi} and T.~{Kitching}, \emph{{Putting the precision in
  precision cosmology: How accurate should your data covariance matrix be?}},
  \href{https://doi.org/10.1093/mnras/stt270}{\emph{"Monthly Notices of the
  Royal Astronomical Society"} {\bfseries 432} (2013) 1928}
  [\href{https://arxiv.org/abs/1212.4359}{{\ttfamily 1212.4359}}].

\bibitem{Plancklensing}
{Planck Collaboration}, N.~{Aghanim}, Y.~{Akrami}, M.~{Ashdown}, J.~{Aumont},
  C.~{Baccigalupi} et~al., \emph{{Planck 2018 results. VIII. Gravitational
  lensing}}, \href{https://doi.org/10.1051/0004-6361/201833886}{\emph{Astron.
  Astrophys.} {\bfseries 641} (2020) A8}
  [\href{https://arxiv.org/abs/1807.06210}{{\ttfamily 1807.06210}}].

\bibitem{SellentinHeavens}
E.~Sellentin and A.F.~Heavens, \emph{{Parameter inference with estimated
  covariance matrices}},
  \href{https://doi.org/10.1093/mnrasl/slv190}{\emph{Monthly Notices of the
  Royal Astronomical Society} {\bfseries 456} (2016) L132}
  [\href{https://arxiv.org/abs/1511.05969}{{\ttfamily 1511.05969}}].

\bibitem{sensitivityanalysis:epidemoilogical}
X.~Lu and E.~Borgonovo, \emph{Global sensitivity analysis in epidemiological
  modeling},
  \href{https://doi.org/https://doi.org/10.1016/j.ejor.2021.11.018}{\emph{European
  Journal of Operational Research} {\bfseries 304} (2023) 9}.

\bibitem{sensitivityanalysis:review}
E.~Borgonovo and E.~Plischke, \emph{Sensitivity analysis: A review of recent
  advances},
  \href{https://doi.org/https://doi.org/10.1016/j.ejor.2015.06.032}{\emph{European
  Journal of Operational Research} {\bfseries 248} (2016) 869}.

\bibitem{sensitivityanalysis:principle}
T.P.~Morris, B.C.~Kahan and I.R.~White, \emph{Choosing sensitivity analyses for
  randomised trials: principles}, {\emph{BMC medical research methodology}
  {\bfseries 14} (2014) 1}.

\bibitem{Chen2003}
G.~Chen, J.R.~Gott, III and B.~Ratra, \emph{{Non-Gaussian error distribution of
  Hubble constant measurements}},
  \href{https://doi.org/10.1086/379219}{\emph{Publ. Astron. Soc. Pac.}
  {\bfseries 115} (2003) 1269}
  [\href{https://arxiv.org/abs/astro-ph/0308099}{{\ttfamily
  astro-ph/0308099}}].

\bibitem{Bailey2017}
D.C.~{Bailey}, \emph{{Not Normal: the uncertainties of scientific
  measurements}}, \href{https://doi.org/10.1098/rsos.160600}{\emph{Royal
  Society Open Science} {\bfseries 4} (2017) 160600}
  [\href{https://arxiv.org/abs/1612.00778}{{\ttfamily 1612.00778}}].

\bibitem{Feeney2018}
S.M.~Feeney, D.J.~Mortlock and N.~Dalmasso, \emph{{Clarifying the Hubble
  constant tension with a Bayesian hierarchical model of the local distance
  ladder}}, \href{https://doi.org/10.1093/mnras/sty418}{\emph{Monthly Notices
  of the Royal Astronomical Society} {\bfseries 476} (2018) 3861}
  [\href{https://arxiv.org/abs/https://academic.oup.com/mnras/article-pdf/476/3/3861/24510928/sty418.pdf}{{\ttfamily
  https://academic.oup.com/mnras/article-pdf/476/3/3861/24510928/sty418.pdf}}].

\bibitem{Bayesiantextbook}
W.M.~Bolstad and J.M.~Curran, \emph{Introduction to Bayesian statistics}, John
  Wiley \& Sons (2016).

\bibitem{Trotta:2017}
R.~Trotta, \emph{{Bayesian Methods in Cosmology}},  1, 2017
  [\href{https://arxiv.org/abs/1701.01467}{{\ttfamily 1701.01467}}].

\bibitem{hartlap2007}
J.~Hartlap, P.~Simon and P.~Schneider, \emph{Why your model parameter
  confidences might be too optimistic. {U}nbiased estimation of the inverse
  covariance matrix},
  \href{https://doi.org/10.1051/0004-6361:20066170}{\emph{Astronomy \&
  Astrophysics} {\bfseries 464} (2007) 399}.

\bibitem{Percival2013}
W.J.~Percival et~al., \emph{{The Clustering of Galaxies in the {SDSS-III}
  Baryon Oscillation Spectroscopic Survey: Including covariance matrix
  errors}}, \href{https://doi.org/10.1093/mnras/stu112}{\emph{Monthly Notices
  of the Royal Astronomical Society} {\bfseries 439} (2014) 2531}
  [\href{https://arxiv.org/abs/1312.4841}{{\ttfamily 1312.4841}}].

\bibitem{York:2000}
D.G.~{York}, J.~{Adelman}, J.~{Anderson}, John~E., S.F.~{Anderson}, J.~{Annis},
  N.A.~{Bahcall} et~al., \emph{{The Sloan Digital Sky Survey: Technical
  Summary}}, \href{https://doi.org/10.1086/301513}{\emph{Astron. J.} {\bfseries
  120} (2000) 1579} [\href{https://arxiv.org/abs/astro-ph/0006396}{{\ttfamily
  astro-ph/0006396}}].

\bibitem{Howlett:2015}
C.~{Howlett}, A.J.~{Ross}, L.~{Samushia}, W.J.~{Percival} and M.~{Manera},
  \emph{{The clustering of the SDSS main galaxy sample - II. Mock galaxy
  catalogues and a measurement of the growth of structure from redshift space
  distortions at z = 0.15}},
  \href{https://doi.org/10.1093/mnras/stu2693}{\emph{Monthly Notices of the
  Royal Astronomical Society} {\bfseries 449} (2015) 848}
  [\href{https://arxiv.org/abs/1409.3238}{{\ttfamily 1409.3238}}].

\bibitem{Ross:2015}
A.J.~{Ross}, L.~{Samushia}, C.~{Howlett}, W.J.~{Percival}, A.~{Burden} and
  M.~{Manera}, \emph{{The clustering of the SDSS DR7 main Galaxy sample - I. A
  4 per cent distance measure at z = 0.15}},
  \href{https://doi.org/10.1093/mnras/stv154}{\emph{Monthly Notices of the
  Royal Astronomical Society} {\bfseries 449} (2015) 835}
  [\href{https://arxiv.org/abs/1409.3242}{{\ttfamily 1409.3242}}].

\bibitem{Abazajian:2009}
{\scshape SDSS Collaboration} collaboration, \emph{{The Seventh Data Release of
  the Sloan Digital Sky Survey}},
  \href{https://doi.org/10.1088/0067-0049/182/2/543}{\emph{Astrophys. J.
  Suppl.} {\bfseries 182} (2009) 543}
  [\href{https://arxiv.org/abs/0812.0649}{{\ttfamily 0812.0649}}].

\bibitem{Dawson:2013}
K.S.~{Dawson}, D.J.~{Schlegel}, C.P.~{Ahn}, S.F.~{Anderson}, {\'E}.~{Aubourg},
  S.~{Bailey} et~al., \emph{{The Baryon Oscillation Spectroscopic Survey of
  SDSS-III}}, \href{https://doi.org/10.1088/0004-6256/145/1/10}{\emph{Astron.
  J.} {\bfseries 145} (2013) 10}
  [\href{https://arxiv.org/abs/1208.0022}{{\ttfamily 1208.0022}}].

\bibitem{Alam-DR11&12:2015}
S.~{Alam}, F.D.~{Albareti}, C.~{Allende Prieto}, F.~{Anders}, S.F.~{Anderson},
  T.~{Anderton} et~al., \emph{{The Eleventh and Twelfth Data Releases of the
  Sloan Digital Sky Survey: Final Data from SDSS-III}},
  \href{https://doi.org/10.1088/0067-0049/219/1/12}{\emph{Astrophys. J. Suppl.}
  {\bfseries 219} (2015) 12}
  [\href{https://arxiv.org/abs/1501.00963}{{\ttfamily 1501.00963}}].

\bibitem{Dawson:2016}
K.S.~{Dawson}, J.-P.~{Kneib}, W.J.~{Percival}, S.~{Alam}, F.D.~{Albareti},
  S.F.~{Anderson} et~al., \emph{{The SDSS-IV Extended Baryon Oscillation
  Spectroscopic Survey: Overview and Early Data}},
  \href{https://doi.org/10.3847/0004-6256/151/2/44}{\emph{Astron. J.}
  {\bfseries 151} (2016) 44}
  [\href{https://arxiv.org/abs/1508.04473}{{\ttfamily 1508.04473}}].

\bibitem{Ahumada:2020}
R.~{Ahumada}, C.A.~{Prieto}, A.~{Almeida}, F.~{Anders}, S.F.~{Anderson},
  B.H.~{Andrews} et~al., \emph{{The 16th Data Release of the Sloan Digital Sky
  Surveys: First Release from the APOGEE-2 Southern Survey and Full Release of
  eBOSS Spectra}},
  \href{https://doi.org/10.3847/1538-4365/ab929e}{\emph{Astrophys. J. Suppl.}
  {\bfseries 249} (2020) 3} [\href{https://arxiv.org/abs/1912.02905}{{\ttfamily
  1912.02905}}].

\bibitem{RossBOSS}
A.J.~Ross, F.~Beutler, C.-H.~Chuang, M.~Pellejero-Ibanez, H.-J.~Seo,
  M.~Vargas-Magaña et~al., \emph{{The clustering of galaxies in the completed
  SDSS-III Baryon Oscillation Spectroscopic Survey: observational systematics
  and baryon acoustic oscillations in the correlation function}},
  \href{https://doi.org/10.1093/mnras/stw2372}{\emph{Monthly Notices of the
  Royal Astronomical Society} {\bfseries 464} (2016) 1168}.

\bibitem{Seo16}
H.-J.~{Seo}, F.~{Beutler}, A.J.~{Ross} and S.~{Saito}, \emph{{Modeling the
  reconstructed BAO in Fourier space}},
  \href{https://doi.org/10.1093/mnras/stw1138}{\emph{Monthly Notices of the
  Royal Astronomical Society} {\bfseries 460} (2016) 2453}
  [\href{https://arxiv.org/abs/1511.00663}{{\ttfamily 1511.00663}}].

\bibitem{Gelman92}
A.~Gelman and D.~Rubin, \emph{Inference from iterative simulation using
  multiple sequences}, {\emph{Statistical Science} {\bfseries 7} (1992) 457}.

\bibitem{HuOkamoto02}
W.~{Hu} and T.~{Okamoto}, \emph{{Mass Reconstruction with Cosmic Microwave
  Background Polarization}},
  \href{https://doi.org/10.1086/341110}{\emph{"Astrophys. J."} {\bfseries 574}
  (2002) 566} [\href{https://arxiv.org/abs/astro-ph/0111606}{{\ttfamily
  astro-ph/0111606}}].

\bibitem{CosmoMC}
A.~Lewis and S.~Bridle, \emph{{Cosmological parameters from {CMB} and other
  data: A {M}onte {C}arlo approach}},
  \href{https://doi.org/10.1103/PhysRevD.66.103511}{\emph{Phys. Rev. D}
  {\bfseries 66} (2002) 103511}
  [\href{https://arxiv.org/abs/astro-ph/0205436}{{\ttfamily
  astro-ph/0205436}}].

\bibitem{Planckresults}
{\scshape Planck} collaboration, \emph{{Planck 2018 results. I. Overview and
  the cosmological legacy of Planck}},
  \href{https://doi.org/10.1051/0004-6361/201833880}{\emph{Astron. Astrophys.}
  {\bfseries 641} (2020) A1}
  [\href{https://arxiv.org/abs/1807.06205}{{\ttfamily 1807.06205}}].

\bibitem{Planck2018-05}
{Planck Collaboration}, N.~{Aghanim}, Y.~{Akrami}, M.~{Ashdown}, J.~{Aumont},
  C.~{Baccigalupi} et~al., \emph{{Planck 2018 results. V. CMB power spectra and
  likelihoods}},
  \href{https://doi.org/10.1051/0004-6361/201936386}{\emph{Astron. Astrophys.}
  {\bfseries 641} (2020) A5}
  [\href{https://arxiv.org/abs/1907.12875}{{\ttfamily 1907.12875}}].

\bibitem{Planck2018}
{\scshape Planck} collaboration, \emph{{Planck 2018 results. VIII.
  Gravitational lensing}},
  \href{https://doi.org/10.1051/0004-6361/201833886}{\emph{Astron. Astrophys.}
  {\bfseries 641} (2020) A8}
  [\href{https://arxiv.org/abs/1807.06210}{{\ttfamily 1807.06210}}].

\bibitem{GBR}
{\O}.~{Rudjord}, N.E.~{Groeneboom}, H.K.~{Eriksen}, G.~{Huey},
  K.M.~{G{\'o}rski} and J.B.~{Jewell}, \emph{{Cosmic Microwave Background
  Likelihood Approximation by a Gaussianized Blackwell-Rao Estimator}},
  \href{https://doi.org/10.1088/0004-637X/692/2/1669}{\emph{Astrophys. J.}
  {\bfseries 692} (2009) 1669}
  [\href{https://arxiv.org/abs/0809.4624}{{\ttfamily 0809.4624}}].

\bibitem{Natale:2020}
U.~Natale, L.~Pagano, M.~Lattanzi, M.~Migliaccio, L.P.~Colombo, A.~Gruppuso
  et~al., \emph{{A novel CMB polarization likelihood package for large angular
  scales built from combined WMAP and Planck LFI legacy maps}},
  \href{https://doi.org/10.1051/0004-6361/202038508}{\emph{Astron. Astrophys.}
  {\bfseries 644} (2020) A32}
  [\href{https://arxiv.org/abs/2005.05600}{{\ttfamily 2005.05600}}].

\bibitem{Beutler}
F.~Beutler, H.-J.~Seo, A.J.~Ross, P.~McDonald, S.~Saito, A.S.~Bolton et~al.,
  \emph{{The clustering of galaxies in the completed SDSS-III Baryon
  Oscillation Spectroscopic Survey: baryon acoustic oscillations in the Fourier
  space}}, \href{https://doi.org/10.1093/mnras/stw2373}{\emph{Monthly Notices
  of the Royal Astronomical Society} {\bfseries 464} (2016) 3409}.

\bibitem{Bautista}
J.E.~Bautista et~al., \emph{{The Completed SDSS-IV extended Baryon Oscillation
  Spectroscopic Survey: measurement of the BAO and growth rate of structure of
  the luminous red galaxy sample from the anisotropic correlation function
  between redshifts 0.6 and 1}},
  \href{https://doi.org/10.1093/mnras/staa2800}{\emph{Monthly Notices of the
  Royal Astronomical Society} {\bfseries 500} (2020) 736}
  [\href{https://arxiv.org/abs/2007.08993}{{\ttfamily 2007.08993}}].

\bibitem{Gil-Marin}
H.~Gil-Marin et~al., \emph{{The Completed SDSS-IV extended Baryon Oscillation
  Spectroscopic Survey: measurement of the BAO and growth rate of structure of
  the luminous red galaxy sample from the anisotropic power spectrum between
  redshifts 0.6 and 1.0}},
  \href{https://doi.org/10.1093/mnras/staa2455}{\emph{Monthly Notices of the
  Royal Astronomical Society} {\bfseries 498} (2020) 2492}
  [\href{https://arxiv.org/abs/2007.08994}{{\ttfamily 2007.08994}}].

\bibitem{Raichoor}
A.~Raichoor et~al., \emph{{The completed SDSS-IV extended Baryon Oscillation
  Spectroscopic Survey: Large-scale Structure Catalogues and Measurement of the
  isotropic BAO between redshift 0.6 and 1.1 for the Emission Line Galaxy
  Sample}}, \href{https://doi.org/10.1093/mnras/staa3336}{\emph{Monthly Notices
  of the Royal Astronomical Society} {\bfseries 500} (2020) 3254}
  [\href{https://arxiv.org/abs/2007.09007}{{\ttfamily 2007.09007}}].

\bibitem{deMattia}
A.~de~Mattia et~al., \emph{{The Completed SDSS-IV extended Baryon Oscillation
  Spectroscopic Survey: measurement of the BAO and growth rate of structure of
  the emission line galaxy sample from the anisotropic power spectrum between
  redshift 0.6 and 1.1}},
  \href{https://doi.org/10.1093/mnras/staa3891}{\emph{Monthly Notices of the
  Royal Astronomical Society} {\bfseries 501} (2021) 5616}
  [\href{https://arxiv.org/abs/2007.09008}{{\ttfamily 2007.09008}}].

\bibitem{Hou}
J.~Hou et~al., \emph{{The Completed SDSS-IV extended Baryon Oscillation
  Spectroscopic Survey: BAO and RSD measurements from anisotropic clustering
  analysis of the Quasar Sample in configuration space between redshift 0.8 and
  2.2}}, \href{https://doi.org/10.1093/mnras/staa3234}{\emph{Monthly Notices of
  the Royal Astronomical Society} {\bfseries 500} (2020) 1201}
  [\href{https://arxiv.org/abs/2007.08998}{{\ttfamily 2007.08998}}].

\bibitem{Neveux}
R.~Neveux et~al., \emph{{The completed SDSS-IV extended Baryon Oscillation
  Spectroscopic Survey: BAO and RSD measurements from the anisotropic power
  spectrum of the quasar sample between redshift 0.8 and 2.2}},
  \href{https://doi.org/10.1093/mnras/staa2780}{\emph{Monthly Notices of the
  Royal Astronomical Society} {\bfseries 499} (2020) 210}
  [\href{https://arxiv.org/abs/2007.08999}{{\ttfamily 2007.08999}}].

\bibitem{EvasCode}
\url{https://github.com/evamariam/CosmoMC_SDSS2020}.

\end{thebibliography}\endgroup

\begin{subappendices}
\renewcommand\thesection{\Alph{section}} 
\renewcommand\thesubsection{\thesection.\arabic{subsection}}
\setcounter{section}{0}

\section{BAO Results: Fitting to Compressed Data} \label{app:BAOcompressed}

\begin{table}[h!]
\centering
\begin{tabular}{|c|c|c|c|c|c|c|}
    \hline
    \multirow{2}{*}{BAO Sample} & $\xi({\bf r})$ & \multirow{2}{*}{\(n_s\)} & \multirow{2}{*}{\(n_d\)} & \multirow{2}{*}{\(n_p\)} & \multirow{2}{*}{Reference} & \multirow{2}{*}{$n_{cp}$} \\
    & or $P({\bf k})$ & & & & & \\
   \hline
   SDSS MGS & $\xi({\bf r})$ & 1000 & 21 & 5 & \multirow{2}{*}{Ross et al. 2015 \cite{Ross:2015}} & \multirow{2}{*}{1} \\
   \cline{2-5}
    DR7 & $P({\bf k})$ & 1000 & 35 & 8 & & \\
   \cline{1-7}
    BOSS LRG & $\xi({\bf r})$ & 1000 & 40 & 10 & Ross et al. 2016 \cite{RossBOSS} & \multirow{2}{*}{4}\\
   \cline{2-6}
    DR12 & $P({\bf k})$ & 1000 & 116 & 16 & Beutler et al. 2016b \cite{Beutler} & \\
   \cline{1-7}
   eBOSS LRG & $\xi({\bf r})$ & 1000 & 40 & 9 & Bautista et al. 2020 \cite{Bautista} & \multirow{2}{*}{2} \\
    \cline{2-6}
    DR16 & $P({\bf k})$ & 1000 & 112 & 17 & Gil-Marin et al. 2020 \cite{Gil-Marin} & \\
    \cline{1-7}
   eBOSS ELG & $\xi({\bf r})$ & 1000 & 40 & 9 &  Raichoor et al. 2020 \cite{Raichoor} & \multirow{2}{*}{1}\\
   \cline{2-6}
    DR16 & $P({\bf k})$ & 1000 & 54 & 13 &  deMattia et al. 2020 \cite{deMattia} & \\
    \cline{1-7}
   eBOSS QSO & $\xi({\bf r})$ & 1000 & 40 & 10 & Hou et al. 2020 \cite{Hou} & \multirow{2}{*}{2} \\
    \cline{2-6}
    DR16 & $P({\bf k})$ & 1000 & 126 & 22 & Neveux et al. 2020 \cite{Neveux} & \\
    \hline
\end{tabular}
\caption{A summary of the methods used to analyse BAO data from various surveys. Here, $\xi({\bf r})$ implies that the correlation function was used as the intermediate statistic, and $P({\bf k})$ implies that the power spectrum was used. The other values stated are the number of simulations (\(n_s\)), the data vector (\(n_d\)), and the number of parameters fitted (\(n_p\)). $n_{cp}$ is the number of cosmological parameters kept after the analysis to be fitted by models, the other $n_p-n_{cp}$ nuisance parameters being marginalised over. All analyses used a Gaussian likelihood with a multiplicative correction as described in \cite{Percival2013}.}
\label{table:covariances}
\end{table}

In this section, rather than re-fitting the correlation function and power spectrum measurements as we did above, we instead work with the publicly released compressed ($\alpha_{\parallel}$, $\alpha_{\perp}$) constraints, which are easier to use and more readily available for a wide range of surveys. This
requires us to make a number of approximations. 
For each data sample, the results from configuration space and Fourier space were combined before being publicly released. In order to allow for this in our revision of the method, we average the values (\(n_s\), \(n_d\), \(n_p\)) given in Table~(\ref{table:covariances}). Following this assumption, we remove the multiplicative factor applied following the derivations of \cite{Percival2013}. We can then convert the cosmological constraints into an estimate of the $\chi^2$ as would be measured from the correlation function and power spectrum for those parameters. 

An additional approximation is required to allow for the ($n_d-n_{cp}$) nuisance parameters that are marginalised over when fitting the 2-point statistics - typically quantifying the broad-band and BAO damping terms. We assume that these do not change the $\chi^2$ values recovered, in part supported by the fact that the multivariate Gaussian and \textit{t}-distributions with the same $\chi^2$ have the same mean values. When evaluating Eq.~\ref{eq:multi-tdist} for the sensitivity tests, we therefore use $n_{cp}$ as the dimension of the scale matrix.

The code \code{CosmoMC} was used to obtain the cosmological parameter constraints for the non-flat o\(\Lambda\)CDM model \cite{CosmoMC}. In particular, we worked with the version of \code{CosmoMC} produced by the SDSS eBOSS collaboration since it contained the setup they used to obtain their results \cite{EvasCode}. We considered two cases, the first where we fit to just the DR12 BOSS LRG $z=0.38$ and $z=0.61$ AP parameters, sampling the parameters \{$\Omega_\mathrm{m}$, $\Omega_\mathrm{k}$\} while fixing \{$\Omega_\mathrm{b}=0.0468$, $H_0=70$\}, matching the analysis done in the main text. We put uninformative, flat priors on $\Omega_\mathrm{m}$ between (0.1, 0.9) and $\Omega_\mathrm{k}$ between (-0.8, 0.8). The second case is where we fit to all BAO measurements and sample the parameters \{$\Omega_\mathrm{m}$, $\Omega_\mathrm{b}$, $H_0$, and $\Omega_\mathrm{k}$\}. In addition to the two priors mentioned already, we also had flat priors on $\Omega_\mathrm{b}$ between (0.001, 0.3) and $H_0$ between (20, 100). 

\subsection{Results for the Likelihood Choice Using Mock-based Covariances}

The o\(\Lambda\)CDM model's cosmological constraints were produced using SDSS's original Gaussian likelihood setup, then with correction factors removed, and followed by applying the Independence-Jeffreys and frequentist-matching prior \textit{t}-distributions. First, we considered fitting just the DR12 BOSS LRG $z=0.38$ and $z=0.61$ AP parameters from \cite{RossBOSS} so that it was the same BAO sample as we use in our main analysis, but with the compressed data rather than correlation functions. The constraints are reported in Table~(\ref{table:BAOcurvBOSSCosmoMC_matchresults}) and show at most a $0.03\sigma$ difference from the Gaussian case. The second case we considered was fitting all of the BAO samples given in Table~(\ref{table:covariances}). These parameter constraints are listed in Table~(\ref{table:BAOcompressed_results}) and show that the results coincide up to 0.05$\sigma$. Both of these results are consistent with the level of sensitivity found in Table~(\ref{table:results}) where we used the full fits. This demonstrates that the choice of likelihood does not have a large effect on the constraints, which holds true when fitting to either correlation functions or compressed data.

\begin{table}[h!]
\centering
\begin{tabular}{|c|c|c|}
    \hline
    \multicolumn{3}{|c|}{BOSS DR12 LRG BAO, $z=0.38,0.61$ } \\
    \hline
     & \(\Omega_\Lambda\) & \(\Omega_\mathrm{k}\) \\
   \hline
    \multirow{2}{*}{Original SDSS method (Gaussian) } & \multirow{2}{*}{\(0.579\pm 0.189\)} & \multirow{2}{*}{\(0.063^{+0.380}_{-0.306}\)} \\
    & &  \\
    \hline
    \multirow{2}{*}{Removed correction factors} & \multirow{2}{*}{\( 0.583\pm 0.185\)} & \multirow{2}{*}{\(0.055^{+0.375}_{-0.300}\)} \\ 
    & &  \\
    \hline
    \multirow{2}{*}{Independence-Jeffreys \textit{t}-dist.} & \multirow{2}{*}{\( 0.584\pm 0.186\)} & \multirow{2}{*}{\(0.055^{+0.371}_{-0.306}\)} \\ 
    & & \\
    \hline
    \multirow{2}{*}{Matching prior \textit{t}-dist.} & \multirow{2}{*}{\( 0.582\pm 0.190\)} & \multirow{2}{*}{\(0.056^{+0.377}_{-0.312}\)} \\ 
    & & \\
    \hline
\end{tabular}
\caption{The posterior mean parameter values and 68\% credible intervals are shown for BOSS DR12 LRG data with $z=0.38,0.61$ and using the o\(\Lambda\)CDM model. We compare results found using the original SDSS method, no correction factors, the Independence-Jeffreys prior $t$-distribution and the matching prior $t$-distribution.}
\label{table:BAOcurvBOSSCosmoMC_matchresults}
\end{table}

\begin{table}[h!]
\centering
\begin{tabular}{|c|c|c|}
    \hline
    \multicolumn{3}{|c|}{All BAO} \\
    \hline
     & \(\Omega_\Lambda\) & \(\Omega_\mathrm{k}\) \\
   \hline 
    \multirow{2}{*}{Original SDSS method (Gaussian)} & \multirow{2}{*}{\(0.507_{-0.104}^{+0.115}\)} & \multirow{2}{*}{\(0.247\pm0.162\)}\\
    & & \\
    \hline
    \multirow{2}{*}{Removed correction factors} & \multirow{2}{*}{\(0.511_{-0.099}^{+0.109}\)} & \multirow{2}{*}{\(0.243_{-0.153}^{+0.154}\)}\\
    & & \\
    \hline
    \multirow{2}{*}{Independence-Jeffreys \textit{t}-dist.} & \multirow{2}{*}{$0.512\pm0.103$} & \multirow{2}{*}{$0.246 _{-0.152}^{+0.154}$}\\
    & & \\
   \hline
    \multirow{2}{*}{Matching prior \textit{t}-dist.} & \multirow{2}{*}{\(0.510_{-0.107}^{+0.107}\)} & \multirow{2}{*}{\(0.248_{-0.158}^{+0.159}\)}\\
    & & \\
    \hline
\end{tabular}
\caption{Parameter constraints for the o\(\Lambda\)CDM model using all BAO measurements. The uncertainties are the 68\%  credible intervals. Results were found using the original SDSS method, no correction factors, then the Independence-Jeffreys prior $t$-distribution and the matching prior $t$-distribution.}
\label{table:BAOcompressed_results}
\end{table}

\subsection{Results from the Sensitivity Analysis}

\begin{table}[h!]
\centering
\begin{tabular}{|c|c|c|}
    \hline
    \multicolumn{3}{|c|}{BOSS DR12 LRG BAO, $z=0.38,0.61$ } \\
    \hline
     & \(\Omega_\Lambda\) & \(\Omega_\mathrm{k}\) \\
   \hline
    \multirow{2}{*}{Original SDSS method (Gaussian)} & \multirow{2}{*}{\(0.579\pm 0.189\)} & \multirow{2}{*}{\(0.063^{+0.380}_{-0.306}\)} \\
    & &  \\
    \hline
    \multirow{2}{*}{\textit{T}-dist. likelihood with $\nu$ = 104 (1\% larger tails) } & \multirow{2}{*}{\(0.582\pm 0.192\)} & \multirow{2}{*}{\(0.056^{+0.368}_{-0.319}\)} \\
    & & \\
    \hline 
    \multirow{2}{*}{\textit{T}-dist. likelihood with $\nu$ = 13 (10\% larger tails)} & \multirow{2}{*}{\(0.581\pm 0.194\)} & \multirow{2}{*}{\(0.057^{+0.374}_{-0.316}\)} \\
    & & \\
    \hline 
\end{tabular}
\caption{Parameter constraints for BOSS DR12 LRG data with $z=0.38,0.61$ and using the o\(\Lambda\)CDM model. The 68\% credible intervals are shown first using SDSS's Gaussian setup and for a \textit{t}-distribution likelihood with 1\% and 10\% more weight in the tails. }
\label{table:BAOcurvBOSSCosmoMC_margresults}
\end{table}

The sensitivity analysis was performed on the compressed BAO data, using both a \textit{t}-distribution with 1\% extra tail probability and one with 10\% extra tail probability. First, the results from fitting just the DR12 BOSS LRG $z=0.38$ and $z=0.61$ AP parameters are shown in Table~(\ref{table:BAOcurvBOSSCosmoMC_margresults}). We can see that the parameter constraints are much less sensitive to the change of likelihood ($\sim 0.02\sigma$) than when we fit to the correlation function data ($\sim 0.2$--$0.4\sigma$), as presented in Table~(\ref{table:BAOmargresults}). This shows that the sensitivity to the form of the likelihood is dependent on fitting directly to the correlation function data, not using the compressed data. We therefore do not present a sensitivity analysis fitting to all BAO samples, as it should be performed on the full fits, not compressed fits.

\end{subappendices}

\end{document}